\documentclass[aps,prb,twocolumn,preprintnumbers,amsmath,amssymb,longbibliography,nofootinbib]{revtex4-2}

\usepackage{graphicx}
\usepackage{dcolumn}
\usepackage{bm}
\usepackage{color}
\usepackage{ulem}
\usepackage[usenames,dvipsnames]{xcolor}
\usepackage{hyperref}
\usepackage{float}

\hypersetup{
    bookmarks=false,        
    colorlinks=true,        
    linkcolor=blue,        
    citecolor=blue,        
    filecolor=blue,      
    urlcolor=blue
}

\newcommand{\be}{\begin{equation}}
\newcommand{\ee}{\end{equation}}
\newcommand{\bea}{\begin{eqnarray}}
\newcommand{\eea}{\end{eqnarray}}

\newcommand{\fig}[1]{Fig.~\ref{#1}}

\newcommand{\e}{\varepsilon}
\newcommand{\w}{\omega}
\newcommand{\s}{\sigma}

\newcommand{\up}{\uparrow}
\newcommand{\down}{\downarrow}

\newcommand{\TKd}{T_K^{SU(4)}}
\newcommand{\TKs}{T_K^{SU(2)}}

\usepackage{ulem}
\usepackage{verbatim} 

\begin{document}

\title{Majorana mode leaking into a spin-charge entangled double quantum dot}

\author{Piotr Majek}
\email{pmajek@amu.edu.pl}
\affiliation{Institute of Spintronics and Quantum Information, Faculty of Physics,
	Adam Mickiewicz University, ul.Uniwersytetu Pozna\'nskiego 2, 61-614 
	Pozna{\'n}, Poland}

\author{Ireneusz Weymann}
\email{weymann@amu.edu.pl}
\affiliation{Institute of Spintronics and Quantum Information, Faculty of Physics,
	Adam Mickiewicz University, ul.Uniwersytetu Pozna\'nskiego 2, 61-614 Pozna{\'n}, Poland}

\date{\today}

\begin{abstract}
The signatures of Majorana zero-energy mode leaking into a spin-charge entangled double quantum dot
are investigated theoretically in the strong electron correlation regime.
The considered setup consists of two capacitively coupled
quantum dots attached to external contacts and side-attached to topological
superconducting wire hosting Majorana quasiparticles.
We show that the presence of Majorana mode gives rise to 
unique features in the local density of states in the $SU(4)$ Kondo regime.
Moreover, it greatly modifies the gate voltage dependence of the linear conductance,
leading to fractional values of the conductance.
We also analyze the effect of a finite length of topological wire
and demonstrate that non-zero overlap of Majorana modes at the ends of the wire
is revealed in local extrema present in the local density of states
of the dot coupled directly to the wire.
The calculations are performed with the aid of the
numerical renormalization group method.
\end{abstract}

\maketitle

\section{Introduction}

One-dimensional topological superconductors or the chains of adatoms on superconducting substrates
are promising platforms to realize Majorana zero-energy modes at the edges \cite{Kitaev2001,Lutchyn2010Aug,Oreg2010Oct,Alicea2012Jun,Kim2018May,Prada2020Oct}.
Signatures of such modes have already been reported by several experiments
\cite{Mourik2012May,Das2012Nov,Deng2012Dec,Churchill2013Jun,Finck2013Mar,
Albrecht2016Mar,Deng2016Dec,jeon_DistinguishingMajorana_2017,nichele_ScalingMajorana_2017,Deng2018Aug,
Lutchyn2018May,Gul2018Jan,zhang_NextSteps_2019}.
There is in fact a great interest in exploring the properties
of such Majorana systems, due to expected applications
in topological quantum computation \cite{kitaev_fault-tolerant_2003,Nayak2008Sep,alicea_NonAbelianStatistics_2011}.
Another important aspect making such hybrid systems interesting
is related to the impact and signatures of the presence of Majorana modes in the
transport properties of attached low-dimensional structures \cite{Liu2011Nov,Leijnse2011Oct,Cao2012Sep,Gong2014Jun,Cheng2014Sep,Liu2015Feb,Deng2016Dec,
	schuray_FanoResonances_2017,prada_MeasuringMajorana_2017,hoffman_SpindependentCoupling_2017,
	gorski_InterplayCorrelations_2018,sanches_MajoranaMolecules_2020,
	wrzesniewski_MagnetizationDynamics_2021}.
In this regard, the leakage of Majorana zero-energy modes into attached 
quantum dots has been a subject of extensive investigations \cite{Vernek2014Apr,Ruiz-Tijerina2015Mar,liu_AndreevBound_2017,zienkiewicz_LeakageMajorana_2019}.
It was shown that the coupling to topological superconductor 
hosting Majorana quasiparticles (Majorana wire) results
in fractional values of the conductance \cite{Lee2013Jun,Vernek2014Apr}.
Moreover, the transport properties of strongly correlated systems
have also been explored in the context of interplay between the Kondo 
and Majorana physics \cite{Golub2011Oct,Lee2013Jun,Cheng2014Sep,eriksson_TunnelingSpectroscopy_2014,
	Weymann2017Apr,Weymann2017Jan,
	Vernek2019,weymann_majorana-kondo_2020,silva_RobustnessKondo_2020}.

\begin{figure}[b]
	\includegraphics[width=1\columnwidth]{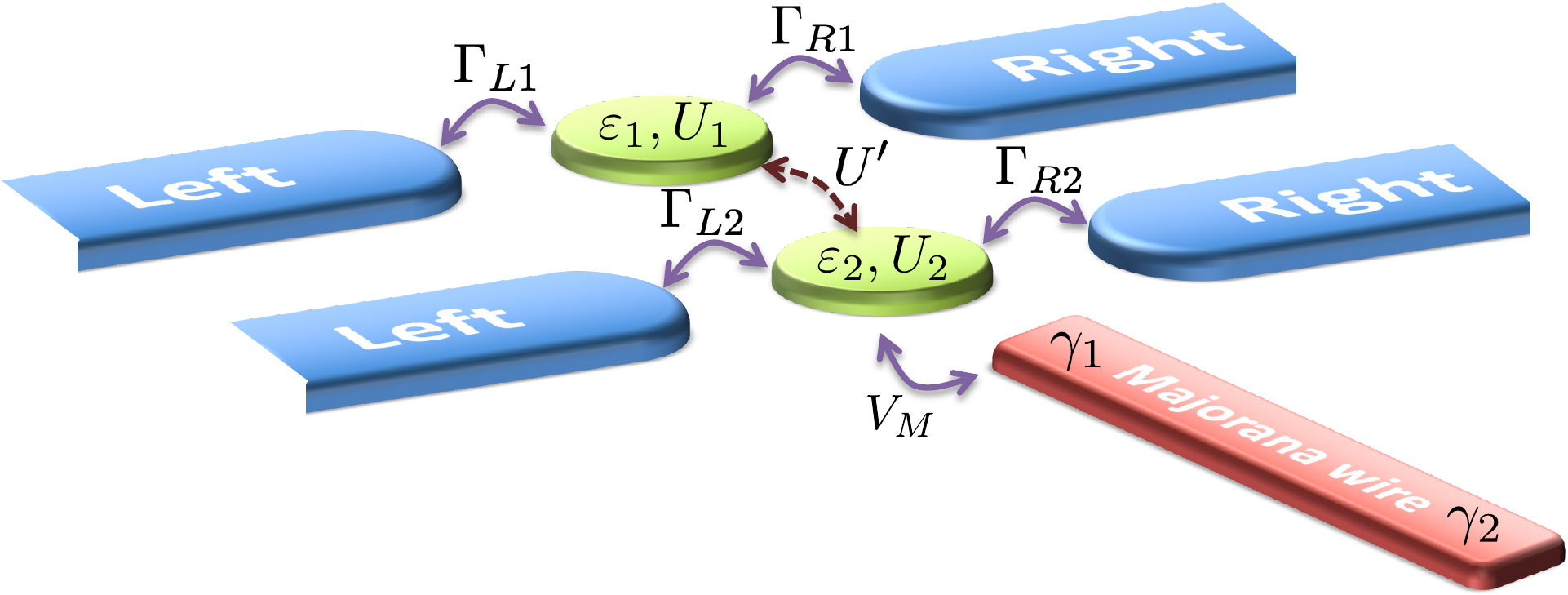}
	\caption{\label{Fig:1}
		The schematic of a double quantum dot
		coupled to a topological superconductor hosting
		Majorana zero-energy modes (Majorana wire).
		Each dot is characterized by its energy $\e_j$
		and Coulomb correlations $U_j$, while $U'$
		denotes the electron correlations between the dots.
		Every dot is connected to separate
		left and right leads with strength $\Gamma_{rj}$.
		The second dot is coupled to the Majorana
		wire (with amplitude $V_M$) hosting Majorana modes described
		by operators $\gamma_1$ and $\gamma_2$.
	}
\end{figure}

Here we make a further step in the understanding of the interplay
between the Majorana zero-energy modes and strong electron correlations in the case
of devices comprising coupled quantum dots.
In particular, we explore the signatures of Majorana quasiparticles in the transport properties
of spin-charge entangled double quantum dot (DQD), see \fig{Fig:1},
exhibiting the $SU(4)$ Kondo effect.
Such Kondo effect results from the four-fold degeneracy
of the ground state due to the orbital and spin degrees of the freedom \cite{hewson_1993,vojtaPRB02,bordaPRL03,Craig2004Science04,SatoP05,
	GalpinPRL05,RuizPRB14,NishikawaPRB16,Lopes2017Jun},
which can be achieved by appropriate tuning of the gate voltages \cite{AmashaPRL13,kellerNP14}.
It is worthy of note that the transport properties of double dots
attached to topological superconducting nanowires
have already been studied \cite{leijnse_ParityQubits_2012,Ivanov2017Jul,cifuentes_ManipulatingMajorana_2019,rancic_EntanglingSpins_2019,
	weymann_majorana-kondo_2020,sanches_MajoranaMolecules_2020,Chen2020Jul}.
However, the system's behavior in the strongly-correlated regime 
is still rather unexplored. The goal of this paper is therefore to shed
more light on this problem. To do so, we employ the nonperturbative
numerical renormalization group (NRG) method \cite{WilsonRMP75},
which allows us to accurately study the equilibrium transport properties of the system.
We determine the relevant spectral functions and
linear conductance through the system for various
strengths of coupling to Majorana wire.
We show that the presence of Majorana zero-energy modes
gives rise to unique features visible in the local density
of states of the double dot. Moreover, we study
the gate voltage dependence of the conductance
and predict its fractional values depending on the 
occupation of the quantum dot system. Finally, we also explore the impact
of finite overlap of Majorana modes and show
that it additionally modifies the behavior
of both the local density of states and the conductance.

The paper is structured as follows. 
Section II is devoted to the theoretical formulation of the problem,
where the Hamiltonian and method are described.
The main results and their discussion are presented
in Sec. III, where we first discuss the behavior of the spectral function
and then analyze the gate voltage and temperature dependence
of conductance. The focus is on both the $SU(4)$ and $SU(2)$ Kondo regimes.
The paper is summarized in Sec. IV.

\section{Theoretical formulation}
\label{sec:modelandmethod}

The system consists of a double quantum dot 
attached to a topological superconducting wire
hosting Majorana zero-energy modes at its ends,
see \fig{Fig:1}.
Each quantum dot is coupled to its own left and right
electrode. The two dots are assumed to be capacitively
coupled, while the hopping between the dots is negligible \cite{kellerNP14}.
The Hamiltonian of the whole system is given by
\be
H = H_{\rm leads} + H_{\rm tun} + H_{\rm MD},
\ee
where the first term describes the 
leads within the free quasiparticle approximation
\be
H_{\rm leads} = \sum_{r=L,R}\sum_{j=1,2} \sum_{\mathbf{k}\sigma}
\e_{rj\mathbf{k}} c^\dag_{rj\mathbf{k}\sigma} c_{rj\mathbf{k}\sigma}.
\ee
Here, $c^\dag_{rj\mathbf{k}\sigma}$ is the creation
operator of a spin-$\sigma$ electron with momentum $\mathbf{k}$
in the left ($r=L$) or right ($r=R$) lead attached to
the first ($j=1$) or second ($j=2$) quantum dot,
and $\e_{rj\mathbf{k}}$ is the corresponding  energy.
The second term of the Hamiltonian describes the
tunneling process between double dot and the leads.
It is given by
\be
H_{\rm tun} = \sum_{r=L,R}\sum_{j=1,2} \sum_{\mathbf{k}\sigma} v_{rj} \left(d^\dag_{j\s}
c_{rj\mathbf{k}\sigma} + c^\dag_{rj\mathbf{k}\sigma} d_{j\s} \right),
\ee
where $d^\dag_{j\s}$ creates an electron with spin $\sigma$
on the $j$th dot and $v_{rj}$ are the tunnel matrix
elements between the dot $j$ and lead $r$.
The quantum dot level broadening is given by
$\Gamma_{rj} = \pi \rho_{rj}v_{rj}^2$, where
$\rho_{rj}$ denotes the corresponding density of states.
Finally, the Hamiltonian of the  double dot attached 
to Majorana wire reads
\bea \label{Eq:HDDM}
H_{\rm MD} &=& \sum_{j\s} \e_j d_{j\s}^\dag d_{j\s}
+ \sum_j U_j d_{j\uparrow}^\dag d_{j\uparrow} d_{j\downarrow}^\dag  d_{j\downarrow} +U' n_1 n_2 
\nonumber\\
&&+ \sqrt{2} V_M (d^\dag_{2\downarrow} \gamma_1 + \gamma_1 d_{2\downarrow}) 
+ i \e_M \gamma_1 \gamma_2,
\eea
where $\e_j$ denotes the energy of an
electron on dot $j$ and $U_j$ stands for the
Coulomb correlation energy between two electrons
occupying the same dot. The Coulomb correlations
between the two quantum dots are denoted by $U'$,
where $n_j = \sum_\s d^\dag_{j\s}d_{j\s}$.
It is assumed that the spin-down component of the second quantum dot is coupled
to the Majorana wire with the amplitude $V_M$
\cite{Flensberg2010Nov,Liu2011Nov,Lee2013Jun,
	Weymann2017Apr,Weymann2017Jan,weymann_majorana-kondo_2020}.
The Majorana operators are denoted by
$\gamma_1$ and $\gamma_2$, respectively,
and $\e_M$ is the overlap between the Majorana  
zero-energy modes \cite{Albrecht2016Mar}.

We are interested in the linear response transport properties
of the system at low temperatures. To accurately resolve this
transport regime, taking into account all electron correlations in a non-perturbative manner,
we make use of the NRG method \cite{WilsonRMP75,BullaRMP08,FlexibleDMNRG}.
This approach allows for a very reliable calculation of various correlation
functions of the system. In particular, we are interested
in the behavior of the spectral function, which can be generally defined as
$A(\omega) = -{\rm Im}\{G^R(\omega)\}/\pi$,  where 
$G^R(\omega)$  is the Fourier transform of the retarded Green's function
$G^R(t) = -i\Theta(t)\langle \{ O^\dag(0), O(t) \} \rangle$,
where $O$ is an operator describing the double dot ($O=d_{j\s}$).

The linear response conductance through the quantum dot $j$ flowing in the spin channel $\sigma$
between the left and right contacts can be found from
\be
	G_{j\sigma} = \frac{e^2}{h}\frac{4\Gamma_{jL}\Gamma_{jR}}{\Gamma_{jL}+\Gamma_{jR}}
	\int d\omega \pi A_{j\sigma}(\omega) \left(-\frac{\partial f(\omega)}{\partial \omega} \right),
\ee
where $f(\omega)$ denotes the Fermi-Dirac distribution function
and $A_{j\sigma}(\omega)$ is the spectral function of quantum dot $j$ for spin $\sigma$.
The conductance through the dot $j$ is then given by
$G_{j} = \sum_\sigma G_{j\s}$, while the total
conductance can be simply expressed as
$G = \sum_j G_j$.

In the following we assume symmetric couplings,
$\Gamma_{rj} = \Gamma/2$ and take $U_1 = U_2\equiv U$.
In NRG calculations we keep at least $N_K=10000$ states
during the iteration and use the band discretization parameter $\Lambda=2$--$2.5$.
Moreover, we make use of the spin and the charge conservation 
of the first dot coupled to its leads. We also exploit the conservation
of spin-up particles and the charge parity for the second dot
coupled to external leads and to the Majorana wire.

\section{Results and discussion}

In this section we present and discuss the numerical results on the transport
properties of the considered double dot-Majorana setup.
We start the considerations with the analysis of the spectral functions.
Then, we study the gate voltage dependence of the linear conductance
for different couplings to topological superconductor.
Finally, we analyze the temperature dependence of the conductance.
We generally focus on the case of a long topological superconducting wire,
such that the overlap between the Majorana zero-energy modes is negligible $\e_M\to 0$.
However, to make the discussion complete, we also analyze the transport behavior
in the short nanowire case, when $\e_M\neq0$.

The considered double dot system exhibits various transport regimes,
where both $SU(4)$ as well as spin or orbital $SU(2)$ Kondo effects can develop
\cite{vojtaPRB02,bordaPRL03,GalpinPRL05,RuizPRB14,Lopes2017Jun,kellerNP14}.
Given the large diversity of the parameter space, in this paper
we focus on the case when the position of the energy level
of each dot is the same, i.e. $\e_1 = \e_2 \equiv \e$.
This condition can be obtained by an appropriate
tuning of the gate voltages \cite{kellerNP14}.
In a such case, in the absence of topological superconducting wire,
the double dot should be 
empty for $\e\gtrsim0$, singly occupied
occupied when $-U'\lesssim \e \lesssim 0$, doubly occupied
with one electron on each dot for $-U'-U\lesssim \e \lesssim -U'$,
occupied by three electrons when $-2U'-U\lesssim \e \lesssim -U'-U$
and the occupation would be full once $\e \lesssim -2U'-U$.
In the transport regime where the DQD occupation is odd,
the system should exhibit the $SU(4)$ Kondo effect.
On the other hand, in the case when the double dot is occupied
with a single electron on each quantum dot
the spin $SU(2)$ Kondo effect develops on every dot.
Below, we thoroughly address the system's transport properties in these two regimes.

\subsection{The spectral functions}
\subsubsection{The $SU(4)$ Kondo regime}

\begin{figure}[t]
	\includegraphics[width=1\columnwidth]{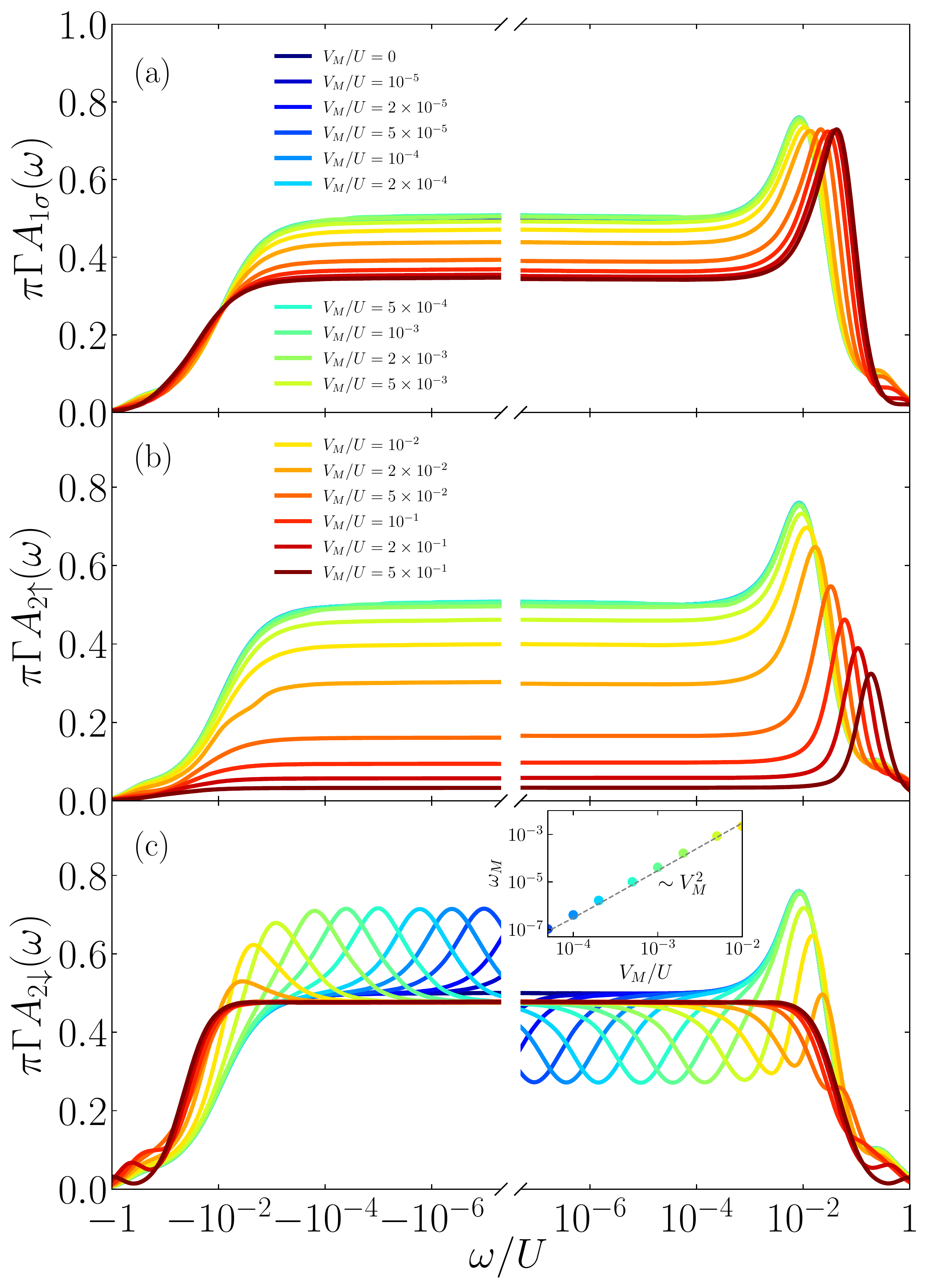} 
	\caption{\label{Fig:A}
		The energy dependence of the normalized 
		spectral function of (a) the first quantum dot, the second quantum dot
		for (b) spin-up and (c) spin-down calculated
		for different values of the coupling to the Majorana wire $V_M$, as indicated.
		The inset in (c) demonstrates the scaling of 
		the position of local minima in $A_{2\down}(\w)$,
		denoted as $\w_M$, with the coupling $V_M$.
		The other parameters are: $U = 1$, $U'=0.5$,
		$\Gamma=0.05$, $\e_1=\e_2=-U'/2$, $T=0$ and $\e_M=0$
		in units of band halfwidth.
	}
\end{figure}

Let us first discuss the behavior of the spectral functions of
the double quantum dot system in the $SU(4)$ Kondo regime.
This Kondo regime is realized when the double dot
is occupied by a single electron and four-fold degeneracy
due to the orbital and spin degrees of freedoms is present,
which happens for $\e_1=\e_2=-U'/2$ \cite{bordaPRL03,kellerNP14}.
The corresponding spectral functions are presented in \fig{Fig:A}
for different values of the coupling to Majorana wire $V_M$.
In the case of $V_M=0$, one can recognize the Kondo resonance
displaced from the Fermi level, a characteristic feature
of the $SU(4)$ Kondo effect. Moreover, one then finds
$A_{1\sigma}(0)  = A_{2\sigma}(0) = 1/2\pi\Gamma$,
which results in $G = \sum_{j\s}G_{j\s}=2e^2/h$ \cite{bordaPRL03,kellerNP14}.
Increasing the coupling to Majorana zero-energy mode
gives rise to a slight decrease of $A_{1\sigma}(0)$
and shift of the Kondo peak to larger energies,
see \fig{Fig:A}(a).

On the other hand, more interesting behavior can
be observed for the spectral function of the second dot,
which is in a direct proximity with topological superconductor.
In the case of spin-up component, one observes
a decrease of $A_{2\up}(0)$ once $V_M\gtrsim 0.005 U \approx \TKd$,
see \fig{Fig:A}(b),
where $\TKd$ is the $SU(4)$ Kondo temperature, the magnitude
of which can be estimated from the displacement 
of the spectral function peak from the Fermi level.
However, now, contrary to the case of the first dot,
the suppression of $A_{2\up}(0)$ is much larger with increasing $V_M$. 
This behavior can be understood by realizing that the coupling
to Majorana wire induces a spin splitting of the dot level,
when the level position is detuned from the particle-hole symmetry point.
Such splitting occurs in the second dot,
and for a single dot proximized by Majorana wire,
but decoupled from normal leads,
it can be expressed as \cite{Lee2013Jun,Weymann2017Apr}
$$\Delta \e_2 = \e_2 + \frac{U}{2} + \frac{1}{2}\sqrt{\e_2^2+V_M^2} 
- \frac{1}{2}\sqrt{(\e_2+U)^2+V_M^2}.$$
In the case of $\e_2 = -U'/2$, by expanding in the leading order in $V_M$,
one obtains
\be \label{Eq:DeltaE2}
\Delta \e_2 \approx \frac{(U-U')V_M^2} {(2U-U')U'}.
\ee
Thus, when $\Delta \e_2 \gtrsim \TKd$, $A_{2\up}(\omega)$
starts depending on $V_M$ and the position of the peak
in the spectral function moves according to $\w \approx \Delta\e_2 \sim V_M^2$.

Because the Majorana mode is assumed to be coupled
to the spin-down level of the second dot, its influence
is most pronounced in $A_{2\down}(\omega)$.
Now, however, one observes new features resulting from the quantum interference
with Majorana mode when the coupling $V_M$ is smaller that the Kondo energy scale.
Interestingly, a local maximum (minimum) develops when $V_M \lesssim \TKd$ 
at negative (positive) energies $|\w|\equiv \w_M$. The minimum moves to larger energies with 
increasing $V_M$ and generates an anti-resonance visible in 
$A_{2\down}(\omega)$ for $\omega \approx \TKd$, until it 
disappears once $V_M \gtrsim \TKd$. On the other hand,
the maximum visible for $\w<0$ moves to larger negative energies and
fades out when the coupling $V_M$ becomes greater than the Kondo temperature.
When this happens, the spectral function exhibits
just a broad plateau with $A_{2\down}(0)=1/2\pi\Gamma$, 
see \fig{Fig:A}(c).
As presented in the inset of \fig{Fig:A}(c),
the position of the local minimum and maximum depends
on the coupling to topological wire as $\w_M \sim V_M^2$.

\begin{figure}[t]
	\includegraphics[width=1\columnwidth]{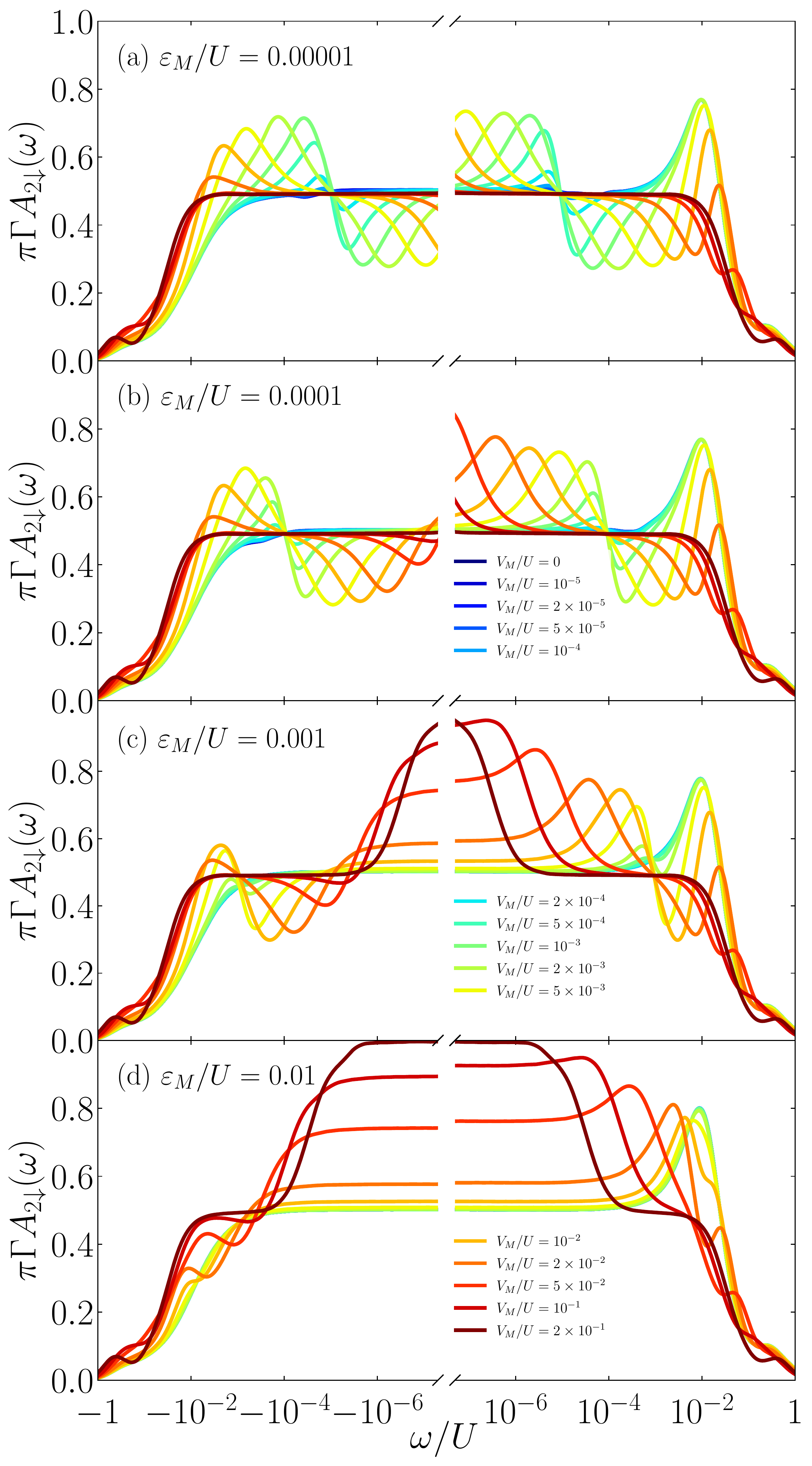} 
	\caption{\label{Fig:AEM}
		The energy dependence of the
		spin-down spectral function of the second quantum dot
		for different values of the coupling to Majorana wire, as indicated,
		and for finite overlap between Majorana modes:
		(a) $\e_M/U = 0.00001$,		
		(b) $\e_M/U = 0.0001$,
		(c) $\e_M/U = 0.001$  and
		(d) $\e_M/U = 0.01$.                              
		The other parameters are the same as in \fig{Fig:A}.
	}
\end{figure}

The effect of finite overlap of Majorana zero-energy modes
on the spectral functions is presented in \fig{Fig:AEM}.
Because this influence is most visible in the case of the spin-down
spectral function of the second dot,
in the figure we present only this correlation function,
while we note that the impact of $\e_M$
on the other spectral functions is very weak
and does not lead to qualitative changes for the considerved values of $\e_M$.
As known from the single quantum dot case \cite{Lee2013Jun,Weymann2017Apr},
the splitting of the Majorana zero-energy modes
destroys the quantum interference and can give rise to the restoration of the
Kondo resonance in the spectral function.
A similar scenario can be observed in the 
case of double quantum dots in the 
spin-charge entangled regime, now, however, the 
dependence on $\e_M$ is much more subtle.
First of all, it can be seen that the energy scale associated with $\e_M$
is clearly visible in the behavior of $A_{2\down}(\omega)$.
For positive (negative) energies the spectral function
displays a minimum (maximum) for $|\omega|\lesssim \TKd$
as long as this local extremum occurs for $|\w|\gtrsim \e_M$,
i.e. as long as $\w_M\gtrsim\e_M$.
On the other hand, for small energies, such that $|\omega|\lesssim \e_M$,
a new local maximum (minimum) develops in $A_{2\down}(\omega)$,
see Figs.~\ref{Fig:AEM}(a)-(c).
This change in the behavior is observed 
when the overlap is smaller than the corresponding Kondo temperature, $\e_M\lesssim \TKd$.
For $\e_M\gtrsim \TKd$, however,
the local extrema are no longer visible,
instead, a pronounced resonance develops at the Fermi energy
for large enough coupling to Majorana wire, see \fig{Fig:AEM}(d).
This resonance signals the orbital Kondo effect,
since the spin-degeneracy is broken by the coupling to topological superconducting wire.

\subsubsection{The $SU(2)$ Kondo regime}

\begin{figure}[t]
	\includegraphics[width=1\columnwidth]{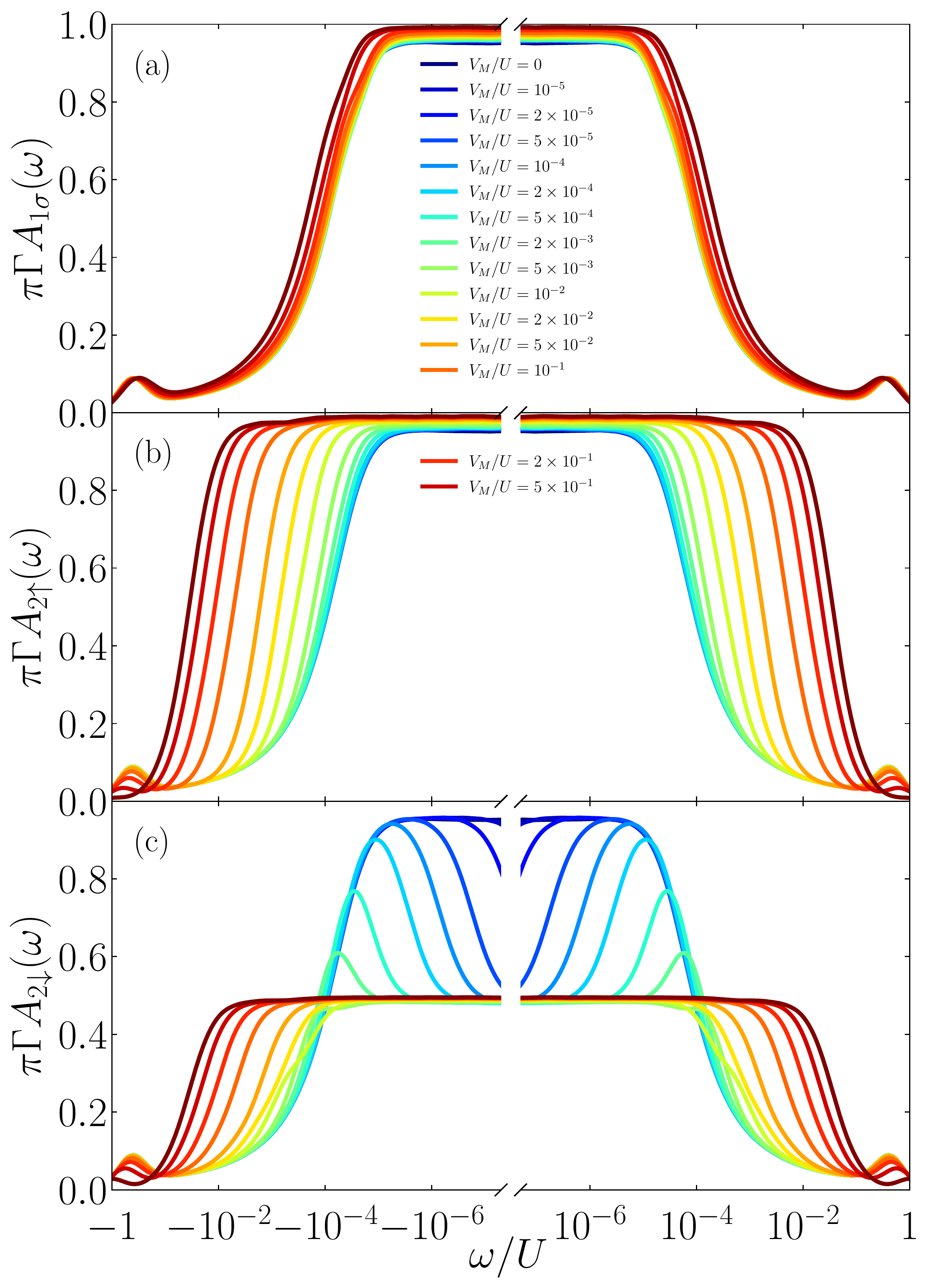} 
	\caption{\label{Fig:ASU2}
		The normalized spectral function of (a) the first quantum dot, the second quantum dot
		for (b) spin-up and (c) spin-down calculated
		for $\e_1 = \e_2 = -U/2-U'$ and for different values of the coupling to the Majorana wire $V_M$, as indicated.
		The other parameters are the same as in \fig{Fig:A}.
	}
\end{figure}

Now we focus on the transport regime where both quantum dots are singly occupied,
such that, in the absence of coupling to Majorana wire,
the spin $SU(2)$ Kondo effect can develop in each dot separately.
The relevant spectral functions in the case of $\e_1 = \e_2 = -U/2-U'$
calculated for different values of $V_M$ are shown in \fig{Fig:ASU2}.
First of all, one can note that the spectral function of the first dot
very weakly depends on $V_M$, while the main impact
is revealed in the behavior of $A_{2\sigma}(\w)$. 
This is quite natural since for $\e = -U/2-U'$ each quantum dot
forms the Kondo state with its own leads and the relevant
energy scale for this state is given by intra-dot correlations
and not inter-dot ones as was in the case of the $SU(4)$ Kondo effect.
Consequently, the coupling to topological superconducting wire
has the main effect on the second dot spectral function.
Moreover, the dependence of $A_{2\sigma}(\w)$ on $V_M$
resembles now that predicted
for the case of a single quantum dot coupled to Majorana wire \cite{Lee2013Jun,Weymann2017Apr}.
One can clearly recognize an energy scale associated with
the coupling to Majorana wire, $\w_M$, which gives rise to 
a local maximum visible in the spin-down spectral function
for low values of $V_M$, see \fig{Fig:ASU2}(c).
Further enhancement of $V_M$, such that $V_M \gtrsim \TKs$,
results in the formation of a plateau at the Fermi energy
of height $A_{2\down}(0) = 1/2\pi\Gamma$,
whose width grows with increasing $V_M$.
This indicates an increase of the associated Kondo temperature,
which can be related to the width of the spectral function resonance at the Fermi level,
see Figs.~\ref{Fig:ASU2}(b) and (c).
It is important to note that, while the coupling to Majorana wire
modifies the Kondo temperature of second dot,
$\TKs$ of the first dot hardly depends on $V_M$.
Consequently, although the double dot 
may still be in the $SU(2)$ Kondo regime,
the strength of Kondo correlations on each dot can be different.

\begin{figure}[t]
	\includegraphics[width=1\columnwidth]{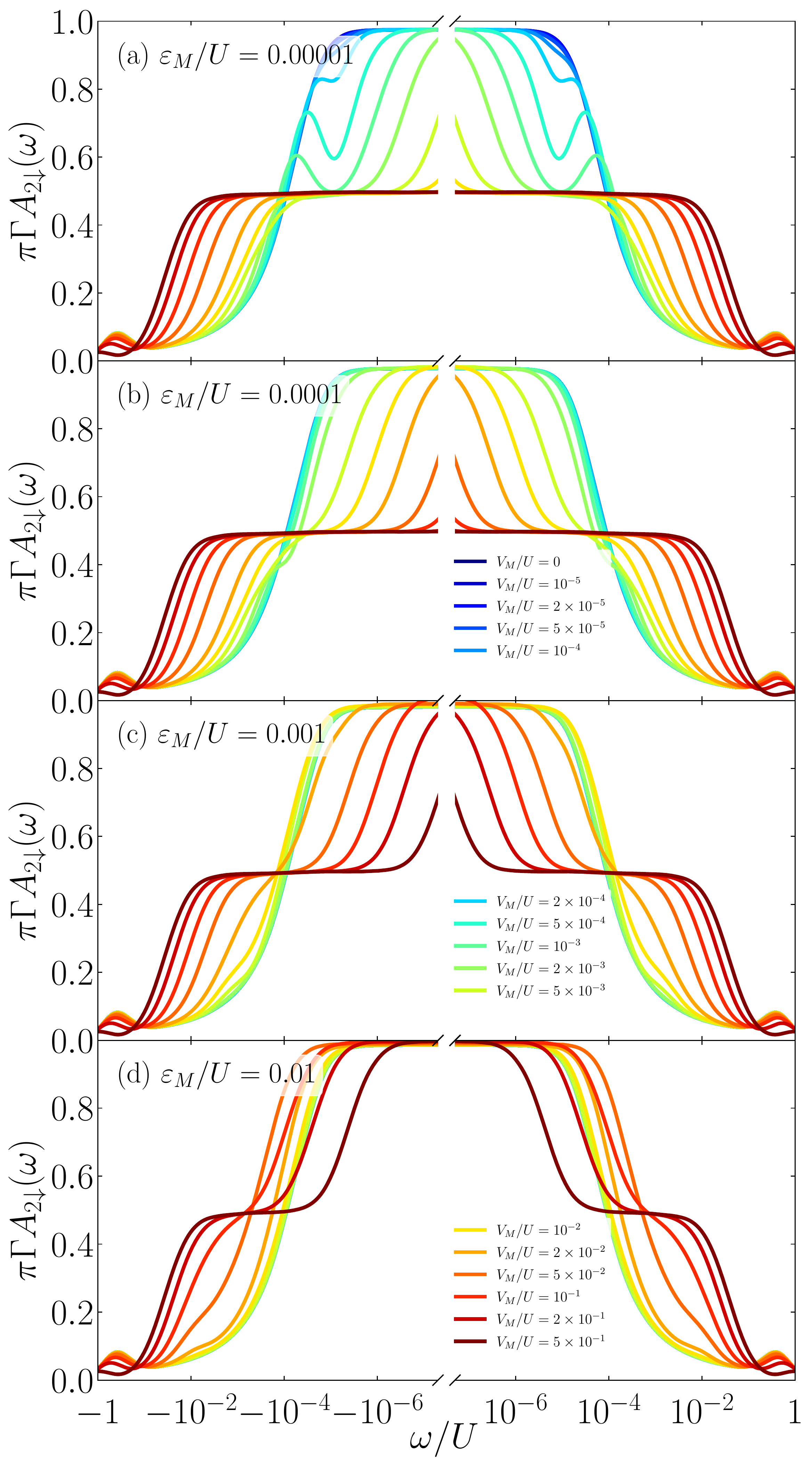} 
	\caption{\label{Fig:ASU2EM}
		The normalized spin-down spectral function of the second dot
		calculated for (a) $\e_M/U=0.00001$, (b)  $\e_M/U=0.0001$, (c)  $\e_M/U=0.001$ 
		and (d)  $\e_M/U=0.01$ for different values of the coupling to the Majorana wire $V_M$, as indicated.
		The other parameters are the same as in \fig{Fig:ASU2}.
	}
\end{figure}

We also analyze the effect of finite length of the Majorana wire on 
the spectral functions in the corresponding transport regime.
As $A_{1\sigma}(\w)$ hardly depends on $V_M$,
it is also independent of $\e_M$.
The dependence of the spin-up spectral function of the second dot on $\e_M$
is mainly quantitative, while no qualitative changes are observed---finite $\e_M$
slightly affects the width of the Kondo resonance, while
its height is not affected. Therefore, in \fig{Fig:ASU2EM}
we only present the dependence of $A_{2\down}(\w)$ on the overlap of Majorana quasiparticles.
When comparing with \fig{Fig:ASU2}(c), one can see that 
finite $\e_M$ suppresses the quantum interference with the Majorana wire
and the local maximum visible in $A_{2\down}(\w)$ for $V_M=0$ at low energies
becomes diminished. Moreover, larger values of $\e_M$
give rise to the restoration of the Kondo peak 
in the spectral function. A similar behavior
has been observed in the case of single quantum dots \cite{Lee2013Jun,Weymann2017Apr}.

\subsection{Gate voltage dependence of conductance}

\begin{figure}[t]
	\includegraphics[width=1\columnwidth]{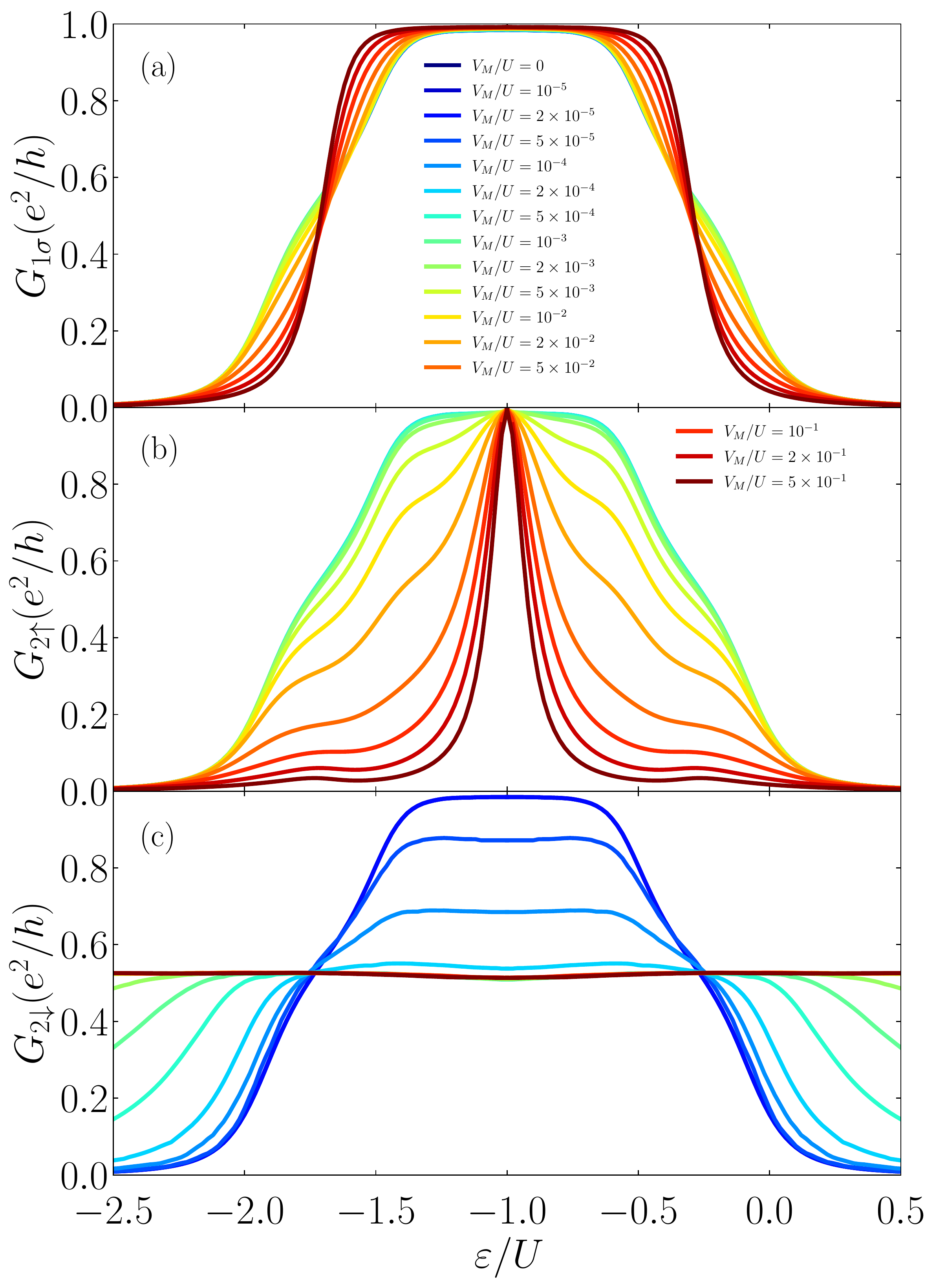} 
	\caption{\label{Fig:G_spin}
		The double dot level dependence of the spin-resolved
		linear conductance of (a) the first quantum dot $G_{1\sigma}$,
		(b) the spin-up $G_{2\up}$ and (c) spin-down $G_{2\down}$ conductance of the second dot.
		The conductance is plotted vs. $\e_1 = \e_2 \equiv \e$
		and calculated for different values of the coupling to the Majorana wire, as indicated.
		The other parameters are the same as in \fig{Fig:A} with $T/U = 5 \cdot 10^{-7}$.
	}
\end{figure}

The spin-resolved linear conductance of both quantum dots
as a function of the double dot energy levels $\e_1 = \e_2 \equiv \e$
and calculated for different values of coupling to Majorana wire
is shown in \fig{Fig:G_spin}. This figure has been calculated
at temperature $T/U = 5 \cdot 10^{-7}$, which is smaller
than the characteristic Kondo scales in the system.
Since the position of the quantum dot levels can be changed
by gate voltages applied to the dots \cite{kellerNP14},
this figure effectively presents the gate voltage dependence of the conductance.

When $V_M=0$ all conductances are equal
and one observes the evolution of $G_{j\sigma}$
with the level position, which is typical for correlated double dots \cite{kellerNP14}.
For $\e\gtrsim 0$, transport
is dominated by elastic cotunneling and the conductance is low.
Then, with lowering $\e$, one enters the $SU(4)$ Kondo regime
and $G_{j\sigma} = e^2/2h$ for $\e = -U'/2$.
When both dots become singly occupied,
which happens when $\e\lesssim -U'$,
the conductance is given by $G_{j\sigma} = e^2/h$
due to the spin $SU(2)$ Kondo effect in each dot.
Note that all the curves are symmetric
with respect to the particle-hole symmetry point of
the double dot system, $\e= -U/2-U'$, see \fig{Fig:G_spin}.

\begin{figure}[t]
	\includegraphics[width=1\columnwidth]{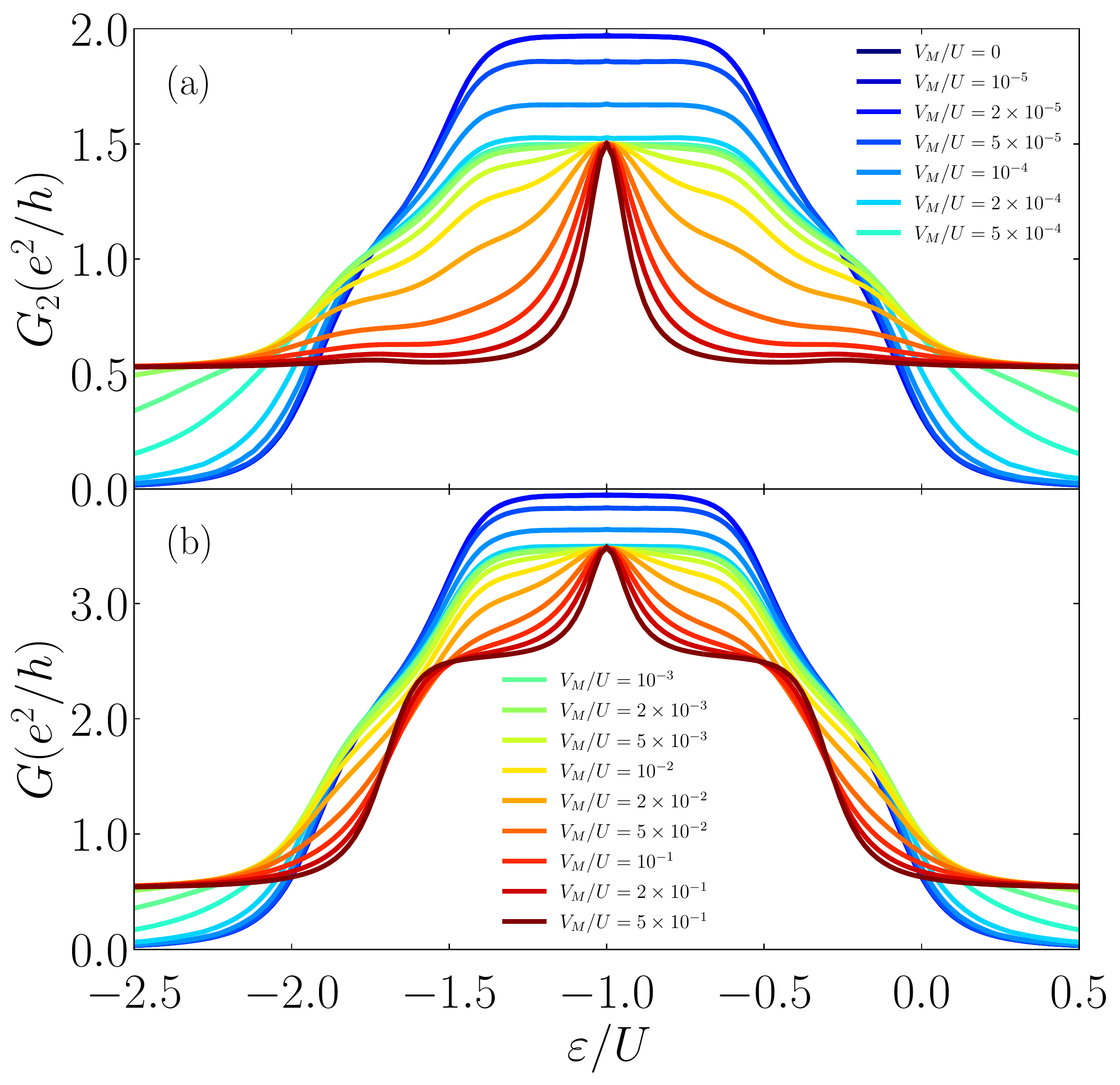} 
	\caption{\label{Fig:G_total}
		The linear conductance of (a) the second dot $G_2$
		and (b) the total conductance $G$ calculated as a function of 
		$\e_1 = \e_2 \equiv \e$ for different values of the coupling to the Majorana wire.
		The parameters are the same as in \fig{Fig:G_spin}.
	}
\end{figure}

When the coupling to Majorana wire is turned on,
the level dependence of conductance becomes modified.
This modification is the largest in the case
of the second dot, while $G_{1\sigma}$ only weakly depends on $V_M$,
cf.  \fig{Fig:G_spin}(a) with Figs.~\ref{Fig:G_spin}(b) and (c).
In particular, the value of $G_{1\sigma} = e^2/h$ around 
the particle-hole symmetry point hardly depends on $V_M$,
while the width of this plateau increases with $V_M$.
In fact, the largest changes of $G_{1\sigma}$
can be observed in the $SU(4)$ Kondo regime
around $\e=-U'/2$, see \fig{Fig:G_spin}(a).
This is just contrary to the behavior of conductance
through the second dot, which strongly depends on the strength
of coupling to topological wire in the full range of $\e$,
see Figs.~\ref{Fig:G_spin}(b) and (c),
except for the particle-hole symmetry point of the double dot,
$\e= -U/2-U'$, in the case of $G_{2\up}$,
where the value of $G_{2\up}$ stays intact as $V_M$ is varied.
Generally, the coupling 
$V_M$ generates a spin-splitting of the second dot level,
which inevitably affects the Kondo state.
For the spin-up conductance component,
this results in the suppression of $G_{2\up}$
except for the particle hole-symmetry point.
On the other hand, for the spin-down component
we observe a mixed behavior: in the $SU(2)$ Kondo regime
the conductance becomes decreased, while in other regimes
it increases with $V_M$ until the conductance eventually reaches $G_{2\down } \approx e^2/2h$,
irrespective of $\e$, once $V_M$ is larger than the
corresponding Kondo scales, 
see Fig.~\ref{Fig:G_spin}(c).
Interestingly, one can also see that with increasing $V_M$
$G_{2\sigma}$ first starts changing in the $SU(2)$ Kondo regime
as compared to the $SU(4)$ Kondo regime. This is due to the
fact that $\TKs < \TKd$, thus, a smaller value of coupling to Majorana
wire is necessary to affect the conductance in the spin Kondo regime.
It is also important to note that for $\e/U\approx -0.25$ 
and $\e/U\approx -1.25$, the conductance $G_{2\down}$
has a stable point in some range of $V_M$, see Fig.~\ref{Fig:G_spin}(c).
This is in fact the middle of the $SU(4)$ Kondo regime
where, because of that, the conductance becomes affected by $V_M$
only when the coupling to Majorana wire is larger than the corresponding Kondo temperature.

Figure \ref{Fig:G_total} presents the full conductance through the second dot,
$G_2 = \sum_\s G_{2\s}$, as well as the total conductance through the system $G$.
Now, one can clearly recognize the influence of the 
presence of topological superconductor hosting Majorana quasiparticles
on the transport behavior of strongly-correlated DQD.
Finite coupling $V_M$ results in fractional values of $G_2$,
namely, $G_2 = 3e^2/2h$ for $\e\approx -U/2-U'$,
while $G_2 = e^2/2h$ otherwise.
In consequence, the total conductance 
in the $SU(2)$ Kondo regime is given by $G = 5e^2/2h$ 
except for $\e\approx -U/2-U'$, where it reaches $G = 7e^2/2h$.
Note that if the coupling to Majorana wire is not large, $V_M/U\lesssim 10^{-3}$,
$G = 7e^2/2h$ in the $SU(2)$ Kondo regime.
On the other hand, for $\e\gtrsim 0$ ($\e\lesssim -U-2U'$)
it becomes $G = e^2/2h$. These fractional
values are clear signatures of the presence of Majorana zero-energy modes in the system.

\begin{figure}[t]
	\includegraphics[width=1\columnwidth]{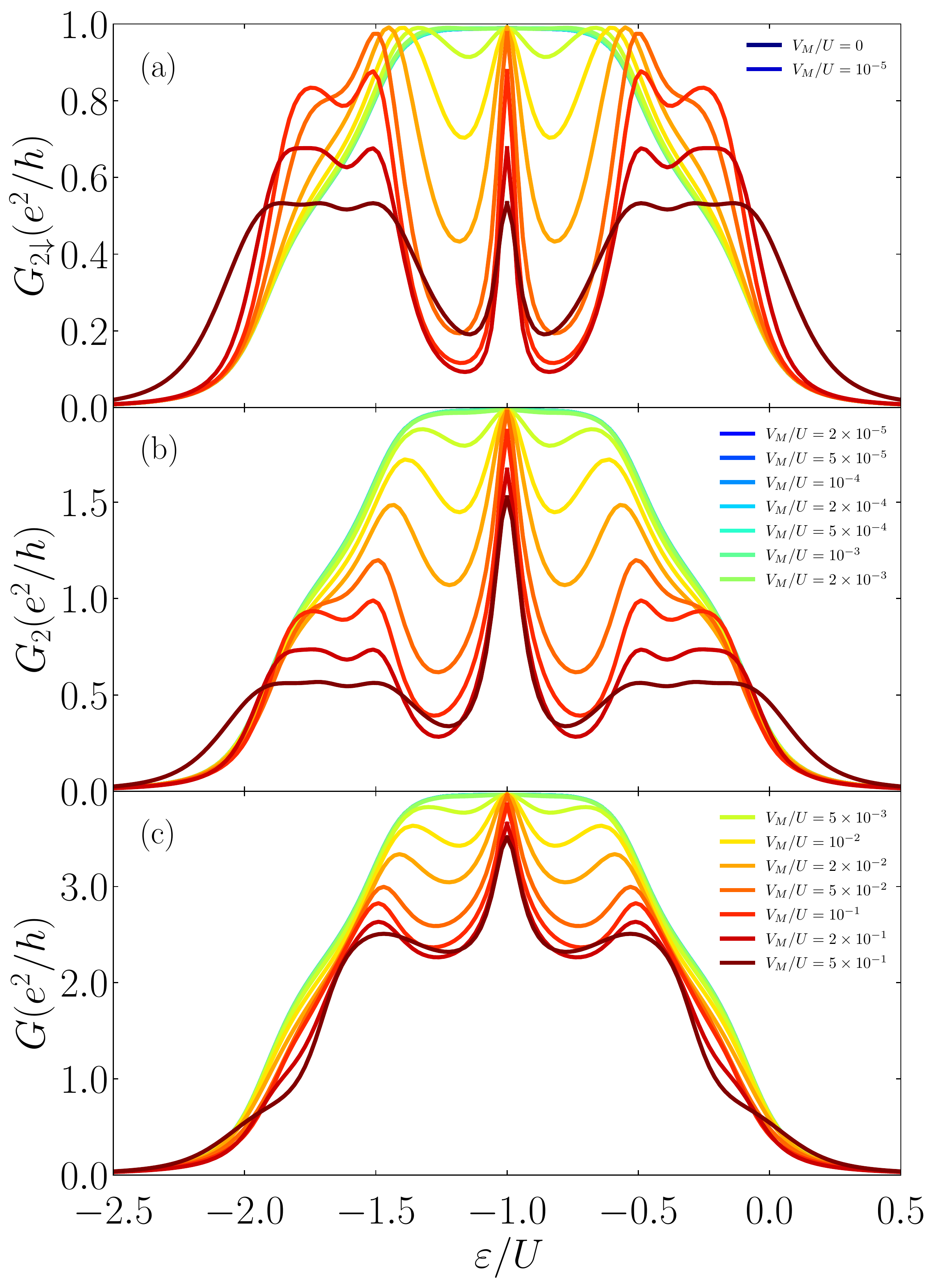} 
	\caption{\label{Fig:GEM}
		The gate voltage dependence of (a) the spin-down
		linear conductance of the second dot,
		(b) the full conductance through the second dot
		as well as (c) the total conductance through the system $G$.
		The conductance is plotted vs. $\e_1 = \e_2 \equiv \e$
		and calculated for different values of the coupling to the Majorana wire,
		as indicated, and for finite overlap between Majorana quasiparticles $\e_M/U=0.001$.
		The other parameters are the same as in \fig{Fig:G_spin}.
	}
\end{figure}

The effect of finite overlap between the Majorana quasiparticles
on the gate voltage dependence of the linear conductance is presented in 
\fig{Fig:GEM}. Because for assumed parameters the impact on both $G_1$ and $G_{2\up}$
is rather weak, we only present the spin-down component of the conductance through the second dot,
together with $G_2$ and $G$, in which the impact of $\e_M$ becomes visible.
Comparing \fig{Fig:GEM}(a) with \fig{Fig:G_spin}(c), one can see that the change
in the behavior is considerable. This is related to the fact that the splitting of Majorana zero-energy modes
suppresses the quantum interference with the topological superconductor.
One can see that when $\e_M>0$, increasing the coupling to the Majorana wire
results in an overall suppression of $G_{2\down}$ in the two-electron transport regime,
i.e. for $-U-U'\lesssim \e \lesssim -U'$, except for the particle-hole symmetry point
$\e=-U/2-U'$. In this transport regime each of the dots exhibits the $SU(2)$ Kondo effect
and the behavior of $G_{2\down}$ is in fact consistent with what has been
observed in the case of single quantum dots coupled to Majorana wire \cite{Weymann2017Apr}.
The coupling to topological superconductor results in a spin splitting
of the dot level, which is somewhat similar to the exchange field induced in the case
of quantum dots attached to ferromagnetic leads \cite{weymannPRB11}. This gives rise
to the suppression of the conductance except for the particle-hole symmetry point.
In the transport regime where the double dot is either empty
or fully occupied, one can see that finite $\e_M$ is responsible for
the suppression of the conductance; $G_{2\down}$ does not saturate 
at $e^2/2h$ with increasing $V_M$ any more, instead it becomes suppressed.
The above described behavior is revealed in the behavior of the 
total conductance through the second dot [\fig{Fig:GEM}(b)]
and the total conductance through the system [\fig{Fig:GEM}(c)].
The main observation is that finite overlap of Majorana modes
changes the fractional values of the conductance present in the case of $\e_M=0$.
The conductance plateau in the middle of the gate voltage dependence
is destroyed and, instead, only a peak with $G_2=2e^2/h$ ($G=4e^2/h$)
for $\e = -U/2=U'$ is present. On the other hand, for $\e\gtrsim 0$
($\e\lesssim -U-2U'$), both $G_2$ and $G$ are generally suppressed.

\subsection{Temperature dependence of conductance}

In this section we present and discuss the effect of coupling to Majorana wire
on the temperature dependence of linear conductance.
We first focus on the case when the system exhibits the $SU(4)$
Kondo effect in the absence of Majorana wire and then
proceed to the case of the spin $SU(2)$ Kondo regime.

\subsubsection{The $SU(4)$ Kondo regime}

\begin{figure}[t]
	\includegraphics[width=1\columnwidth]{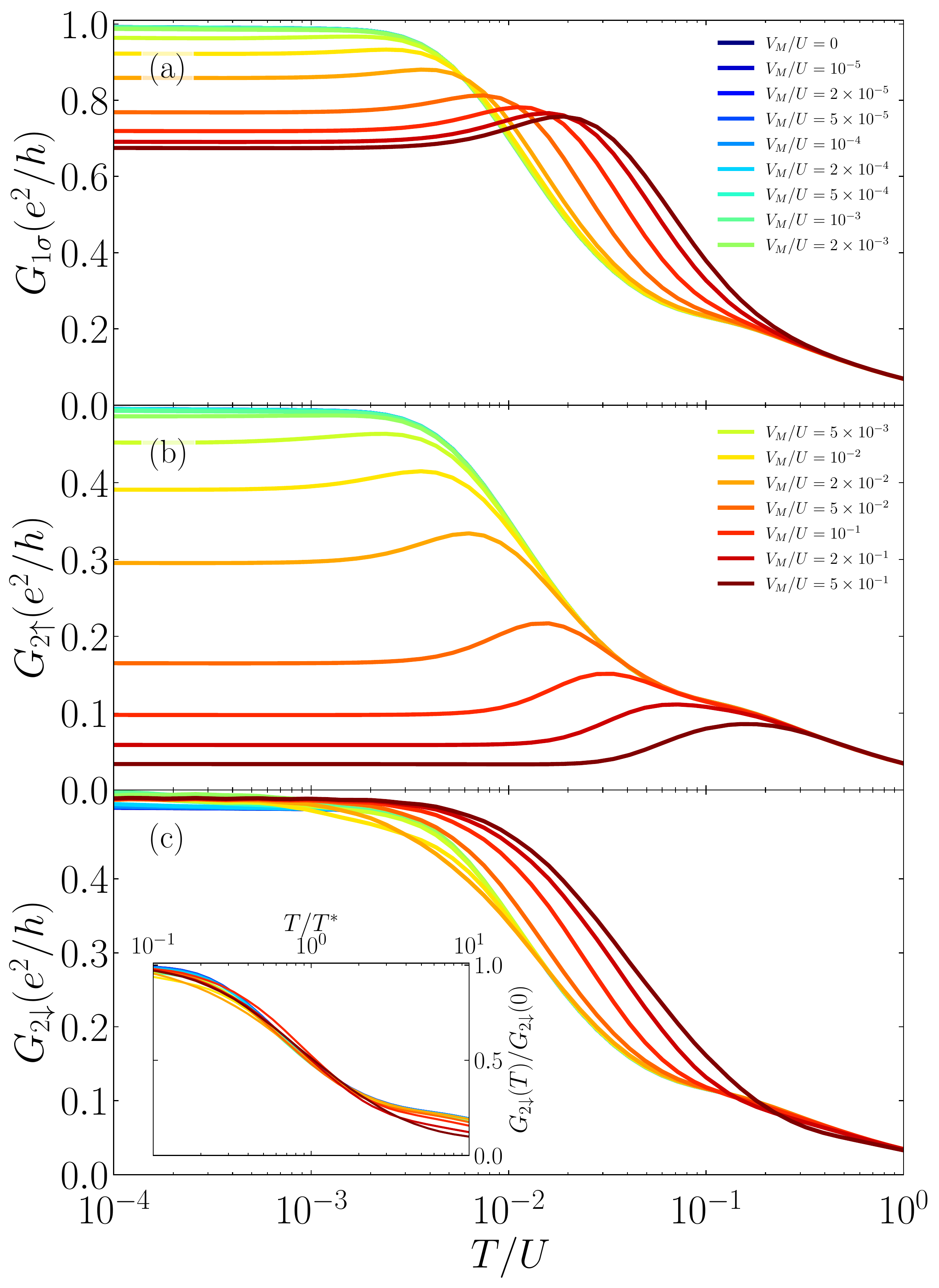} 
	\caption{\label{Fig:GT_spin}
		The temperature dependence of the linear conductance of (a) the first 
		quantum dot, the second quantum dot for (b) spin-up and (c) spin-down
		calculated for $\e_1=\e_2=-U'/2$ and different values of the coupling to the Majorana wire, as indicated.
		The inset in (c) presents the normalized conductance
		$G_{2\down}(T)/G_{2\down}(0)$
		as a function of $T/T^*$, where $T^*$ is the temperature at which
		$G_{2\down}(T)/G_{2\down}(0) = 1/2$.
		The other parameters are the same as in \fig{Fig:A}.
	}
\end{figure}

The quantum dot- and spin-resolved conductance as a function
of temperature calculated for $\e_1=\e_2 =-U'/2$ is displayed in \fig{Fig:GT_spin}.
For $V_M=0$, the conductance displays 
a scaling behavior characteristic of the $SU(4)$ Kondo regime.
When the coupling to the Majorana wire is turned on,
with its grow, one observes a gradual distortion of the universal conductance curve.
When $V_M$ is sufficiently large, $G_1$ exhibits a small drop
of its low-temperature value, while a local maximum develops around
$T\approx \TKd$, which moves to higher temperatures with increasing $V_M$,
see \fig{Fig:GT_spin}(a).
Much larger influence is visible in the case of the second dot,
where almost a full suppression of the conductance is visible
in the spin-up channel, with a local maximum at
the temperatures corresponding to a similar behavior present in $G_1$ 
[\fig{Fig:GT_spin}(b)].
On the other hand, the spin-down channel of the second dot
reveals a weaker dependence on $V_M$.
The low-temperature maximum of $G_{2\down}$ basically does not depend
on $V_M$, whereas the temperature at which the conductance reaches maximum
slightly increases with $V_M$. 
In the inset of \fig{Fig:GT_spin}(c)
we present explicitly the scaling behavior of the conductance $G_{2\down}$.
The normalized conductance $G_{2\down}(T)/G_{2\down}(0)$
is plotted vs $T/T^*$, where $T^*$ is the temperature at which
the conductance drops to a half of its low-temperature value.
Consequently, for $V_M$, one has $T^* = \TKd$,
which yields $\TKd/U \approx 0.019$.
One can see that for finite $V_M$ 
the temperature dependence of the normalized conductance becomes modified
and it deviates from the $SU(4)$ scaling.

\begin{figure}[t]
	\includegraphics[width=1\columnwidth]{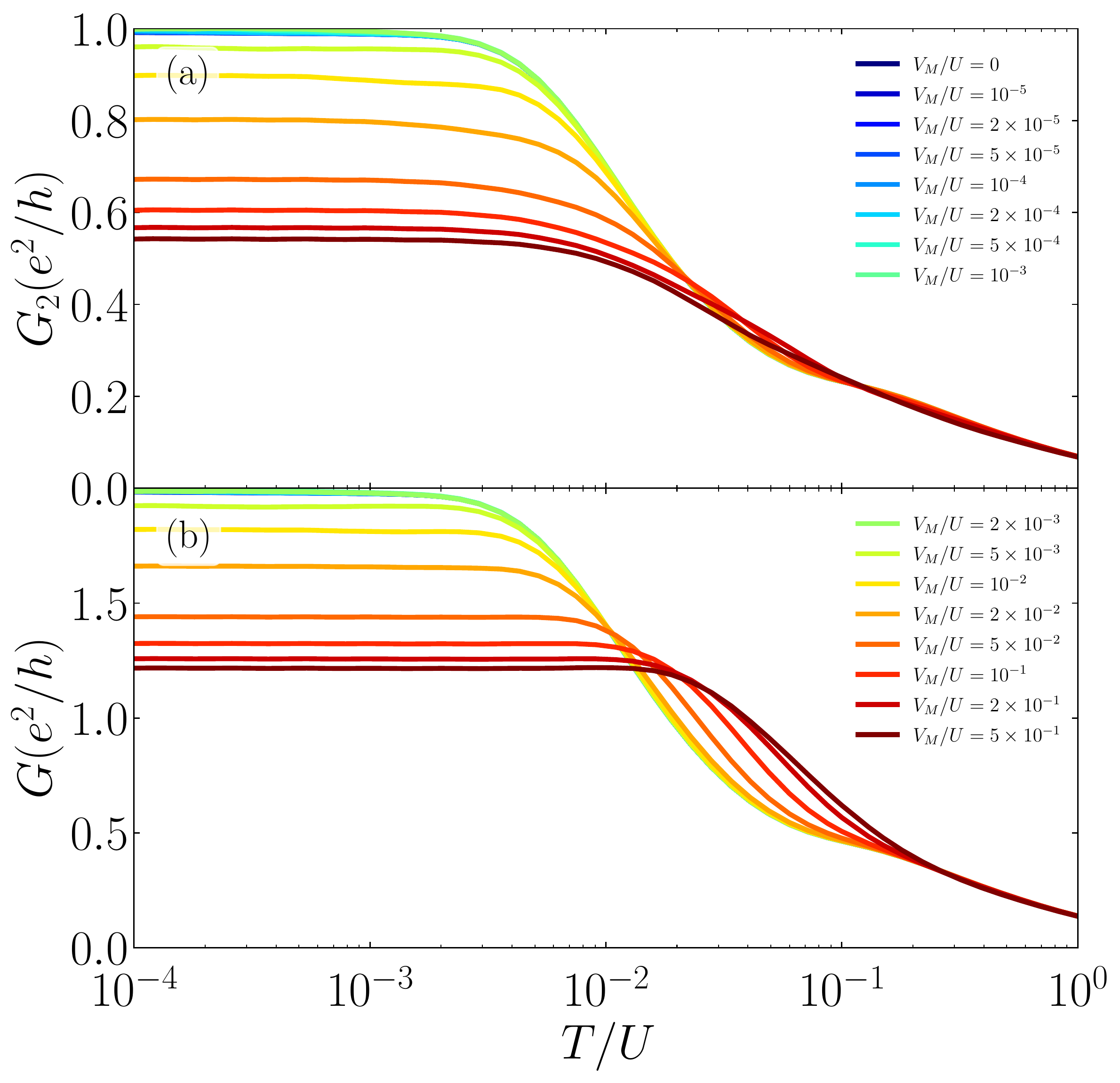} 
	\caption{\label{Fig:GT_total}
		The temperature dependence of the linear conductance of (a) 
		the second quantum dot, (b) and the total conductance of the system,
		calculated for different values of the coupling to the Majorana wire.
		The parameters are the same as in \fig{Fig:GT_spin}.
	}
\end{figure}

The conductance through the second dot and
the total conductance of the system as a function of temperature
are shown in \fig{Fig:GT_total}. 
One can now clearly see that increasing $V_M$ results
in lowering of the low-temperature conductance.
Moreover, a slight increase of the temperature
at which the conductance starts raising due to the Kondo effect
is also visible. 

\subsubsection{The $SU(2)$ Kondo regime}

\begin{figure}[t]
	\includegraphics[width=1\columnwidth]{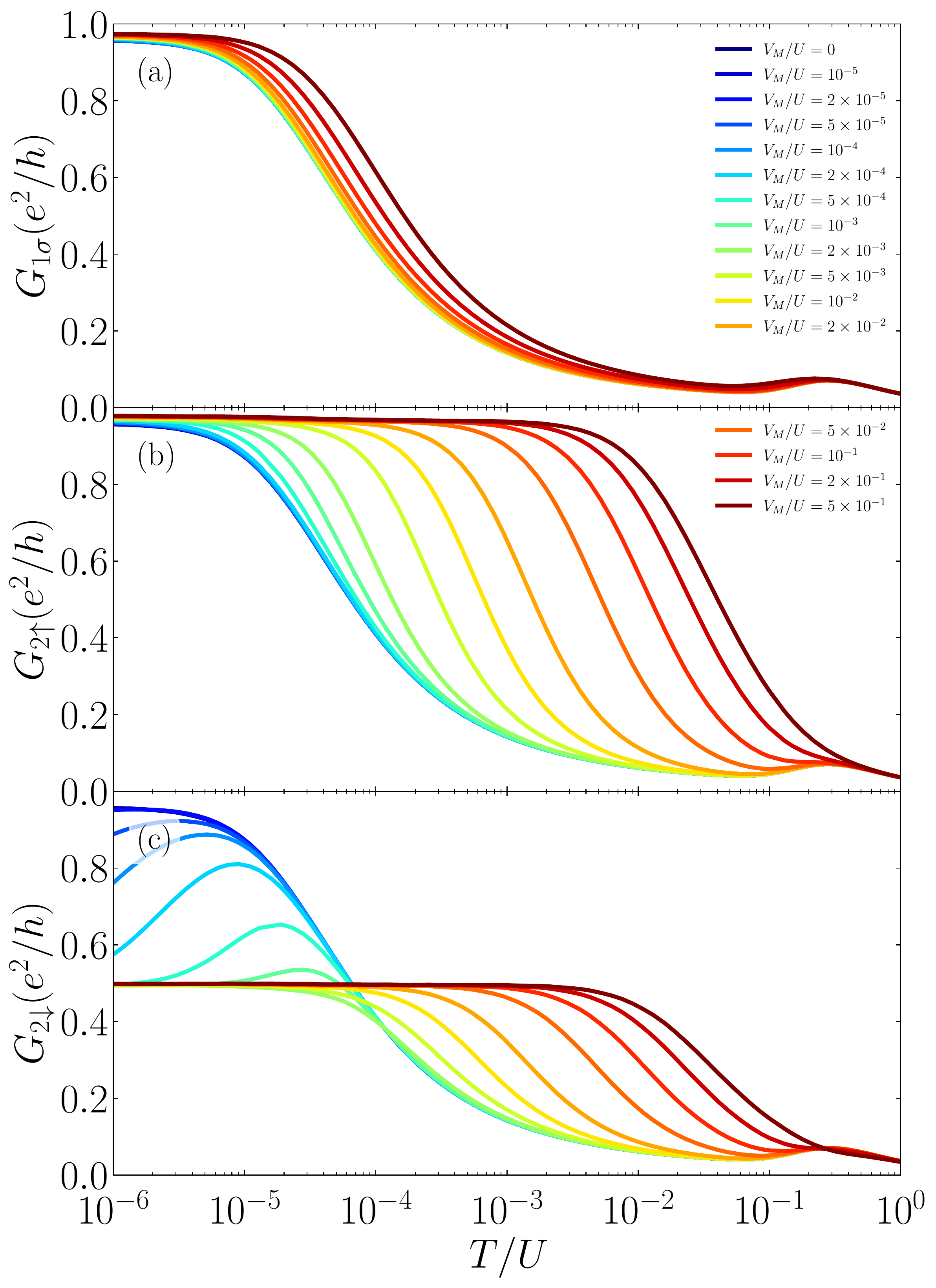} 
	\caption{\label{Fig:GTSU2}
		The temperature dependence of the linear conductance of (a) first 
		quantum dot, the second quantum dot for (b) spin-up and (c) spin-down
		calculated for $\e_1=\e_2=-U/2-U'$ and for different values of the coupling to the Majorana wire, as indicated.
		The other parameters are the same as in \fig{Fig:ASU2}.
	}
\end{figure}

Figure~\ref{Fig:GTSU2} presents the temperature dependence
of the spin-resolved conductance through each dot
calculated for different temperatures and assuming $\e=-U/2-U'$.
When $V_M=0$, the system exhibits the Kondo effect
on every quantum dot and the conductance reaches its maximum value,
i.e. $G_{j\sigma} = e^2/h$. When the coupling to Majorana wire is turned on,
it very weakly affects the temperature dependence of $G_{1\sigma}$,
while the largest changes can be seen in the behavior of $G_{2\sigma}$.
More specifically, in the case of the first dot one only observes a small
increase of the corresponding Kondo temperature.
This is contrary to $G_{2\up}$, which exhibits a strong enhancement
of the Kondo temperature with increasing $V_M$, see Fig.~\ref{Fig:GTSU2}(b).
A similar enhancement of energy scale at which conductance increases
can be observed in the case of  $G_{2\down}$ shown in Fig.~\ref{Fig:GTSU2}(c).
Now, however, one can see a nonmonotonic behavior of
conductance for small values of $V_M$ and consecutive suppression
of the low-temperature conductance to $G_{2\down} = e^2/2h$.

\begin{figure}[t]
	\includegraphics[width=1\columnwidth]{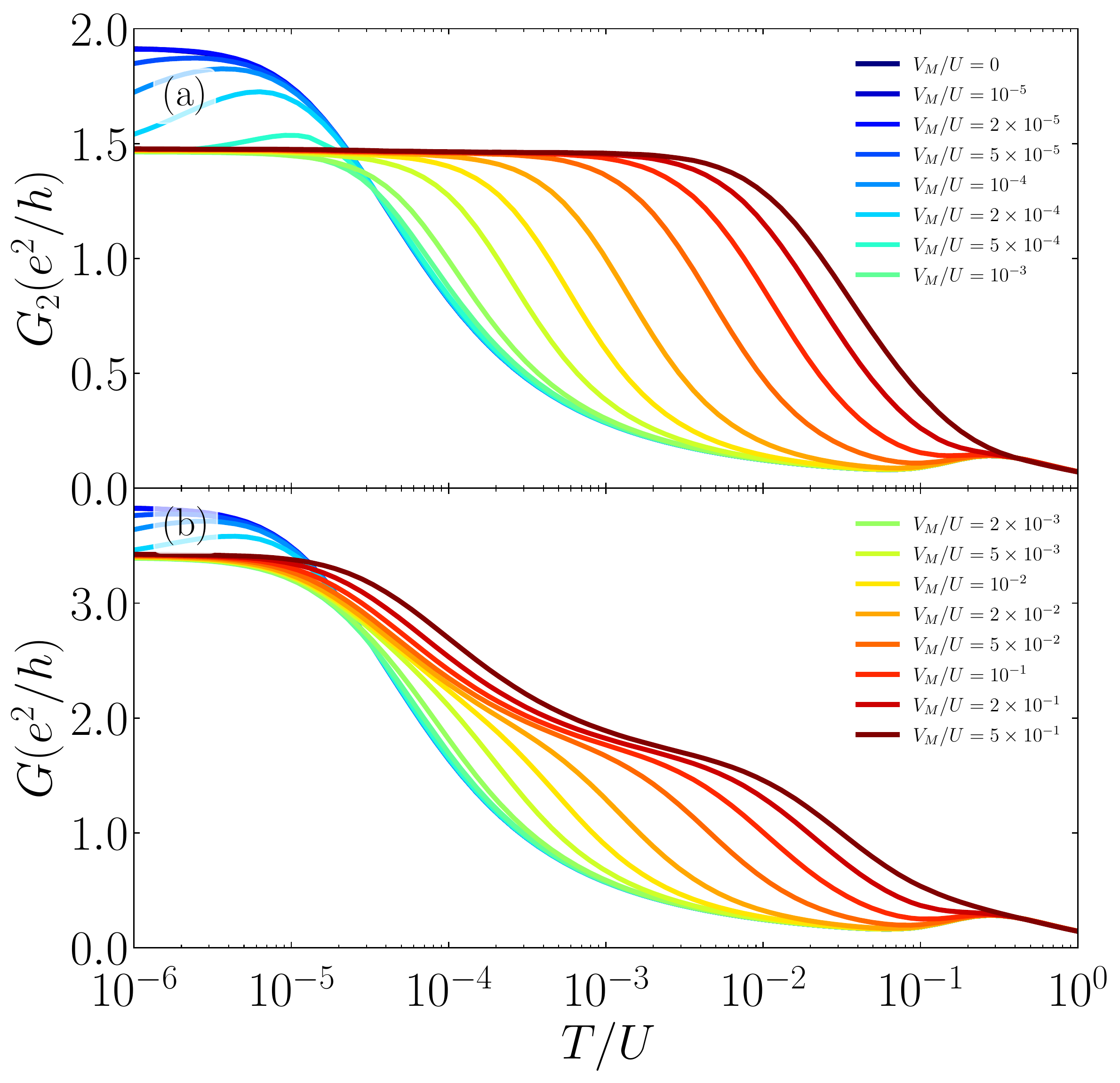} 
	\caption{\label{Fig:GTSU2f}
		The temperature dependence of (a) the total conductance 
		through the second dot and (b) the total conductance through the system
		for different values of $V_M$.
		The parameters are the same as in \fig{Fig:GTSU2}.
	}
\end{figure}

To make the picture complete, in \fig{Fig:GTSU2f}
we show the total conductance through the second dot and 
the total conductance through the system.
The temperature dependence of $G_2$ shows a characteristic
dip at low temperatures and low values of $V_M$,
resulting from the quantum interference with the Majorana wire.
This behavior bears some resemblance to the single quantum dot case \cite{Weymann2017Apr}.
On the other hand, the total conductance for larger values
of coupling to topological superconductor exhibits a two-step
increase with lowering the temperature.
First, $G$ reaches the value of the order of $3e^2/2h$
and then, with further decrease of $T$, it approaches
$G = 7e^2/2h$, see \fig{Fig:GTSU2f}(b).

\section{Conclusions}

We have analyzed the transport properties
of strongly correlated double quantum dot coupled to the Majorana wire.
The calculations have been performed with the aid of the 
numerical renormalization group method.
The main focus was on the signatures and impact of Majorana quasiparticles
on spin-charge entangled Kondo regime of the system.
However, we have also addressed the transport regime
where the system exhibits the spin Kondo effect.
In particular, we have comprehensively analyzed 
the spectral functions of the dots as well as
the gate voltage and temperature dependence of the linear conductance.
We have considered the case of a long Majorana wire,
however, the effect of finite overlap of Majorana zero-energy modes
has also been addressed.

We have shown that in the $SU(4)$ Kondo regime
the spectral function of the dot in the vicinity of topological superconductor
exhibits a local minimum (maximum) on either side of the Fermi level.
Moreover, in the case of finite overlap between Majorana modes,
a local maximum and minimum may occur
both for positive and negative energies, indicating the energy
scale associated with the splitting of Majorana quasiparticles.
In the Coulomb valley with a single electron on each dot,
the coupling to Majorana wire results in a local minimum at low energies visible in the 
spin-down spectral function of the second dot. 
This feature becomes however suppressed when
the Majorana wire is relatively short.

The behavior of the spectral functions is directly revealed in the
linear conductance. Studying the gate voltage
dependence of spin-resolved conductance, we have
identified transport regimes where $G$ reaches fractional values,
which stem from the presence of Majorana quasiparticles in the system.
We have also demonstrated that these fractional conductance features
may become suppressed if a considerable overlap between Majorana modes is present.
Finally, we have determined the temperature dependence of the conductance
shedding light on its behavior both in the spin-charge entangled regime
as well as in the transport regime where the system exhibits the spin $SU(2)$
Kondo effect. Deviations from universal scaling characteristic of those
regimes are uncovered.

We believe that our study extends the knowledge on the interplay
of Kondo correlations with topological superconducting wires hosting Majorana modes,
fostering further efforts to study and understand the physics
of two fundamental research areas:
topological properties of matter and nontrivial Kondo states.

\begin{acknowledgments}
This work was supported by the National Science Centre
in Poland through the Project No. 2018/29/B/ST3/00937.
The computing time at the Pozna\'n Supercomputing
and Networking Center is acknowledged.
\end{acknowledgments}



\begin{thebibliography}{67}%
	\makeatletter
	\providecommand \@ifxundefined [1]{%
		\@ifx{#1\undefined}
	}%
	\providecommand \@ifnum [1]{%
		\ifnum #1\expandafter \@firstoftwo
		\else \expandafter \@secondoftwo
		\fi
	}%
	\providecommand \@ifx [1]{%
		\ifx #1\expandafter \@firstoftwo
		\else \expandafter \@secondoftwo
		\fi
	}%
	\providecommand \natexlab [1]{#1}%
	\providecommand \enquote  [1]{``#1''}%
	\providecommand \bibnamefont  [1]{#1}%
	\providecommand \bibfnamefont [1]{#1}%
	\providecommand \citenamefont [1]{#1}%
	\providecommand \href@noop [0]{\@secondoftwo}%
	\providecommand \href [0]{\begingroup \@sanitize@url \@href}%
	\providecommand \@href[1]{\@@startlink{#1}\@@href}%
	\providecommand \@@href[1]{\endgroup#1\@@endlink}%
	\providecommand \@sanitize@url [0]{\catcode `\\12\catcode `\$12\catcode
		`\&12\catcode `\#12\catcode `\^12\catcode `\_12\catcode `\%12\relax}%
	\providecommand \@@startlink[1]{}%
	\providecommand \@@endlink[0]{}%
	\providecommand \url  [0]{\begingroup\@sanitize@url \@url }%
	\providecommand \@url [1]{\endgroup\@href {#1}{\urlprefix }}%
	\providecommand \urlprefix  [0]{URL }%
	\providecommand \Eprint [0]{\href }%
	\providecommand \doibase [0]{https://doi.org/}%
	\providecommand \selectlanguage [0]{\@gobble}%
	\providecommand \bibinfo  [0]{\@secondoftwo}%
	\providecommand \bibfield  [0]{\@secondoftwo}%
	\providecommand \translation [1]{[#1]}%
	\providecommand \BibitemOpen [0]{}%
	\providecommand \bibitemStop [0]{}%
	\providecommand \bibitemNoStop [0]{.\EOS\space}%
	\providecommand \EOS [0]{\spacefactor3000\relax}%
	\providecommand \BibitemShut  [1]{\csname bibitem#1\endcsname}%
	\let\auto@bib@innerbib\@empty
	\bibitem [{\citenamefont {Kitaev}(2001)}]{Kitaev2001}%
	\BibitemOpen
	\bibfield  {author} {\bibinfo {author} {\bibfnamefont {A.~Y.}\ \bibnamefont
			{Kitaev}},\ }\bibfield  {title} {\bibinfo {title} {{Unpaired Majorana
				fermions in quantum wires}},\ }\href
	{https://doi.org/10.1070/1063-7869/44/10s/s29} {\bibfield  {journal}
		{\bibinfo  {journal} {Phys. Usp.}\ }\textbf {\bibinfo {volume} {44}},\
		\bibinfo {pages} {131} (\bibinfo {year} {2001})}\BibitemShut {NoStop}%
	\bibitem [{\citenamefont {Lutchyn}\ \emph {et~al.}(2010)\citenamefont
		{Lutchyn}, \citenamefont {Sau},\ and\ \citenamefont
		{Das~Sarma}}]{Lutchyn2010Aug}%
	\BibitemOpen
	\bibfield  {author} {\bibinfo {author} {\bibfnamefont {R.~M.}\ \bibnamefont
			{Lutchyn}}, \bibinfo {author} {\bibfnamefont {J.~D.}\ \bibnamefont {Sau}},\
		and\ \bibinfo {author} {\bibfnamefont {S.}~\bibnamefont {Das~Sarma}},\
	}\bibfield  {title} {\bibinfo {title} {{Majorana Fermions and a Topological
				Phase Transition in Semiconductor-Superconductor Heterostructures}},\ }\href
	{https://doi.org/10.1103/PhysRevLett.105.077001} {\bibfield  {journal}
		{\bibinfo  {journal} {Phys. Rev. Lett.}\ }\textbf {\bibinfo {volume} {105}},\
		\bibinfo {pages} {077001} (\bibinfo {year} {2010})}\BibitemShut {NoStop}%
	\bibitem [{\citenamefont {Oreg}\ \emph {et~al.}(2010)\citenamefont {Oreg},
		\citenamefont {Refael},\ and\ \citenamefont {von Oppen}}]{Oreg2010Oct}%
	\BibitemOpen
	\bibfield  {author} {\bibinfo {author} {\bibfnamefont {Y.}~\bibnamefont
			{Oreg}}, \bibinfo {author} {\bibfnamefont {G.}~\bibnamefont {Refael}},\ and\
		\bibinfo {author} {\bibfnamefont {F.}~\bibnamefont {von Oppen}},\ }\bibfield
	{title} {\bibinfo {title} {{Helical Liquids and Majorana Bound States in
				Quantum Wires}},\ }\href {https://doi.org/10.1103/PhysRevLett.105.177002}
	{\bibfield  {journal} {\bibinfo  {journal} {Phys. Rev. Lett.}\ }\textbf
		{\bibinfo {volume} {105}},\ \bibinfo {pages} {177002} (\bibinfo {year}
		{2010})}\BibitemShut {NoStop}%
	\bibitem [{\citenamefont {Alicea}(2012)}]{Alicea2012Jun}%
	\BibitemOpen
	\bibfield  {author} {\bibinfo {author} {\bibfnamefont {J.}~\bibnamefont
			{Alicea}},\ }\bibfield  {title} {\bibinfo {title} {{New directions in the
				pursuit of Majorana fermions in solid state systems}},\ }\href
	{https://doi.org/10.1088/0034-4885/75/7/076501} {\bibfield  {journal}
		{\bibinfo  {journal} {Rep. Prog. Phys.}\ }\textbf {\bibinfo {volume} {75}},\
		\bibinfo {pages} {076501} (\bibinfo {year} {2012})}\BibitemShut {NoStop}%
	\bibitem [{\citenamefont {Kim}\ \emph {et~al.}(2018)\citenamefont {Kim},
		\citenamefont {Palacio-Morales}, \citenamefont {Posske}, \citenamefont
		{R{\ifmmode\acute{o}\else\'{o}\fi}zsa}, \citenamefont
		{Palot{\ifmmode\acute{a}\else\'{a}\fi}s}, \citenamefont {Szunyogh},
		\citenamefont {Thorwart},\ and\ \citenamefont {Wiesendanger}}]{Kim2018May}%
	\BibitemOpen
	\bibfield  {author} {\bibinfo {author} {\bibfnamefont {H.}~\bibnamefont
			{Kim}}, \bibinfo {author} {\bibfnamefont {A.}~\bibnamefont
			{Palacio-Morales}}, \bibinfo {author} {\bibfnamefont {T.}~\bibnamefont
			{Posske}}, \bibinfo {author} {\bibfnamefont {L.}~\bibnamefont
			{R{\ifmmode\acute{o}\else\'{o}\fi}zsa}}, \bibinfo {author} {\bibfnamefont
			{K.}~\bibnamefont {Palot{\ifmmode\acute{a}\else\'{a}\fi}s}}, \bibinfo
		{author} {\bibfnamefont {L.}~\bibnamefont {Szunyogh}}, \bibinfo {author}
		{\bibfnamefont {M.}~\bibnamefont {Thorwart}},\ and\ \bibinfo {author}
		{\bibfnamefont {R.}~\bibnamefont {Wiesendanger}},\ }\bibfield  {title}
	{\bibinfo {title} {{Toward tailoring Majorana bound states in artificially
				constructed magnetic atom chains on elemental superconductors}},\ }\href
	{https://doi.org/10.1126/sciadv.aar5251} {\bibfield  {journal} {\bibinfo
			{journal} {Sci. Adv.}\ }\textbf {\bibinfo {volume} {4}},\ \bibinfo {pages}
		{eaar5251} (\bibinfo {year} {2018})}\BibitemShut {NoStop}%
	\bibitem [{\citenamefont {Prada}\ \emph {et~al.}(2020)\citenamefont {Prada},
		\citenamefont {San-Jose}, \citenamefont {de~Moor}, \citenamefont {Geresdi},
		\citenamefont {Lee}, \citenamefont {Klinovaja}, \citenamefont {Loss},
		\citenamefont {Nyg{\aa}rd}, \citenamefont {Aguado},\ and\ \citenamefont
		{Kouwenhoven}}]{Prada2020Oct}%
	\BibitemOpen
	\bibfield  {author} {\bibinfo {author} {\bibfnamefont {E.}~\bibnamefont
			{Prada}}, \bibinfo {author} {\bibfnamefont {P.}~\bibnamefont {San-Jose}},
		\bibinfo {author} {\bibfnamefont {M.~W.~A.}\ \bibnamefont {de~Moor}},
		\bibinfo {author} {\bibfnamefont {A.}~\bibnamefont {Geresdi}}, \bibinfo
		{author} {\bibfnamefont {E.~J.~H.}\ \bibnamefont {Lee}}, \bibinfo {author}
		{\bibfnamefont {J.}~\bibnamefont {Klinovaja}}, \bibinfo {author}
		{\bibfnamefont {D.}~\bibnamefont {Loss}}, \bibinfo {author} {\bibfnamefont
			{J.}~\bibnamefont {Nyg{\aa}rd}}, \bibinfo {author} {\bibfnamefont
			{R.}~\bibnamefont {Aguado}},\ and\ \bibinfo {author} {\bibfnamefont {L.~P.}\
			\bibnamefont {Kouwenhoven}},\ }\bibfield  {title} {\bibinfo {title} {{From
				Andreev to Majorana bound states in hybrid
				superconductor{\textendash}semiconductor nanowires}},\ }\href
	{https://doi.org/10.1038/s42254-020-0228-y} {\bibfield  {journal} {\bibinfo
			{journal} {Nat. Rev. Phys.}\ }\textbf {\bibinfo {volume} {2}},\ \bibinfo
		{pages} {575} (\bibinfo {year} {2020})}\BibitemShut {NoStop}%
	\bibitem [{\citenamefont {Mourik}\ \emph {et~al.}(2012)\citenamefont {Mourik},
		\citenamefont {Zuo}, \citenamefont {Frolov}, \citenamefont {Plissard},
		\citenamefont {Bakkers},\ and\ \citenamefont {Kouwenhoven}}]{Mourik2012May}%
	\BibitemOpen
	\bibfield  {author} {\bibinfo {author} {\bibfnamefont {V.}~\bibnamefont
			{Mourik}}, \bibinfo {author} {\bibfnamefont {K.}~\bibnamefont {Zuo}},
		\bibinfo {author} {\bibfnamefont {S.~M.}\ \bibnamefont {Frolov}}, \bibinfo
		{author} {\bibfnamefont {S.~R.}\ \bibnamefont {Plissard}}, \bibinfo {author}
		{\bibfnamefont {E.~P. A.~M.}\ \bibnamefont {Bakkers}},\ and\ \bibinfo
		{author} {\bibfnamefont {L.~P.}\ \bibnamefont {Kouwenhoven}},\ }\bibfield
	{title} {\bibinfo {title} {{Signatures of Majorana Fermions in Hybrid
				Superconductor-Semiconductor Nanowire Devices}},\ }\href
	{https://doi.org/10.1126/science.1222360} {\bibfield  {journal} {\bibinfo
			{journal} {Science}\ }\textbf {\bibinfo {volume} {336}},\ \bibinfo {pages}
		{1003} (\bibinfo {year} {2012})}\BibitemShut {NoStop}%
	\bibitem [{\citenamefont {Das}\ \emph {et~al.}(2012)\citenamefont {Das},
		\citenamefont {Ronen}, \citenamefont {Most}, \citenamefont {Oreg},
		\citenamefont {Heiblum},\ and\ \citenamefont {Shtrikman}}]{Das2012Nov}%
	\BibitemOpen
	\bibfield  {author} {\bibinfo {author} {\bibfnamefont {A.}~\bibnamefont
			{Das}}, \bibinfo {author} {\bibfnamefont {Y.}~\bibnamefont {Ronen}}, \bibinfo
		{author} {\bibfnamefont {Y.}~\bibnamefont {Most}}, \bibinfo {author}
		{\bibfnamefont {Y.}~\bibnamefont {Oreg}}, \bibinfo {author} {\bibfnamefont
			{M.}~\bibnamefont {Heiblum}},\ and\ \bibinfo {author} {\bibfnamefont
			{H.}~\bibnamefont {Shtrikman}},\ }\bibfield  {title} {\bibinfo {title}
		{{Zero-bias peaks and splitting in an Al{\textendash}InAs nanowire
				topological superconductor as a signature of Majorana fermions}},\ }\href
	{https://doi.org/10.1038/nphys2479} {\bibfield  {journal} {\bibinfo
			{journal} {Nat. Phys.}\ }\textbf {\bibinfo {volume} {8}},\ \bibinfo {pages}
		{887} (\bibinfo {year} {2012})}\BibitemShut {NoStop}%
	\bibitem [{\citenamefont {Deng}\ \emph {et~al.}(2012)\citenamefont {Deng},
		\citenamefont {Yu}, \citenamefont {Huang}, \citenamefont {Larsson},
		\citenamefont {Caroff},\ and\ \citenamefont {Xu}}]{Deng2012Dec}%
	\BibitemOpen
	\bibfield  {author} {\bibinfo {author} {\bibfnamefont {M.~T.}\ \bibnamefont
			{Deng}}, \bibinfo {author} {\bibfnamefont {C.~L.}\ \bibnamefont {Yu}},
		\bibinfo {author} {\bibfnamefont {G.~Y.}\ \bibnamefont {Huang}}, \bibinfo
		{author} {\bibfnamefont {M.}~\bibnamefont {Larsson}}, \bibinfo {author}
		{\bibfnamefont {P.}~\bibnamefont {Caroff}},\ and\ \bibinfo {author}
		{\bibfnamefont {H.~Q.}\ \bibnamefont {Xu}},\ }\bibfield  {title} {\bibinfo
		{title} {{Anomalous Zero-Bias Conductance Peak in a Nb{\textendash}InSb
				Nanowire{\textendash}Nb Hybrid Device}},\ }\href
	{https://doi.org/10.1021/nl303758w} {\bibfield  {journal} {\bibinfo
			{journal} {Nano Lett.}\ }\textbf {\bibinfo {volume} {12}},\ \bibinfo {pages}
		{6414} (\bibinfo {year} {2012})}\BibitemShut {NoStop}%
	\bibitem [{\citenamefont {Churchill}\ \emph {et~al.}(2013)\citenamefont
		{Churchill}, \citenamefont {Fatemi}, \citenamefont {Grove-Rasmussen},
		\citenamefont {Deng}, \citenamefont {Caroff}, \citenamefont {Xu},\ and\
		\citenamefont {Marcus}}]{Churchill2013Jun}%
	\BibitemOpen
	\bibfield  {author} {\bibinfo {author} {\bibfnamefont {H.~O.~H.}\
			\bibnamefont {Churchill}}, \bibinfo {author} {\bibfnamefont {V.}~\bibnamefont
			{Fatemi}}, \bibinfo {author} {\bibfnamefont {K.}~\bibnamefont
			{Grove-Rasmussen}}, \bibinfo {author} {\bibfnamefont {M.~T.}\ \bibnamefont
			{Deng}}, \bibinfo {author} {\bibfnamefont {P.}~\bibnamefont {Caroff}},
		\bibinfo {author} {\bibfnamefont {H.~Q.}\ \bibnamefont {Xu}},\ and\ \bibinfo
		{author} {\bibfnamefont {C.~M.}\ \bibnamefont {Marcus}},\ }\bibfield  {title}
	{\bibinfo {title} {{Superconductor-nanowire devices from tunneling to the
				multichannel regime: Zero-bias oscillations and magnetoconductance
				crossover}},\ }\href {https://doi.org/10.1103/PhysRevB.87.241401} {\bibfield
		{journal} {\bibinfo  {journal} {Phys. Rev. B}\ }\textbf {\bibinfo {volume}
			{87}},\ \bibinfo {pages} {241401(R)} (\bibinfo {year} {2013})}\BibitemShut
	{NoStop}%
	\bibitem [{\citenamefont {Finck}\ \emph {et~al.}(2013)\citenamefont {Finck},
		\citenamefont {Van~Harlingen}, \citenamefont {Mohseni}, \citenamefont
		{Jung},\ and\ \citenamefont {Li}}]{Finck2013Mar}%
	\BibitemOpen
	\bibfield  {author} {\bibinfo {author} {\bibfnamefont {A.~D.~K.}\
			\bibnamefont {Finck}}, \bibinfo {author} {\bibfnamefont {D.~J.}\ \bibnamefont
			{Van~Harlingen}}, \bibinfo {author} {\bibfnamefont {P.~K.}\ \bibnamefont
			{Mohseni}}, \bibinfo {author} {\bibfnamefont {K.}~\bibnamefont {Jung}},\ and\
		\bibinfo {author} {\bibfnamefont {X.}~\bibnamefont {Li}},\ }\bibfield
	{title} {\bibinfo {title} {{Anomalous Modulation of a Zero-Bias Peak in a
				Hybrid Nanowire-Superconductor Device}},\ }\href
	{https://doi.org/10.1103/PhysRevLett.110.126406} {\bibfield  {journal}
		{\bibinfo  {journal} {Phys. Rev. Lett.}\ }\textbf {\bibinfo {volume} {110}},\
		\bibinfo {pages} {126406} (\bibinfo {year} {2013})}\BibitemShut {NoStop}%
	\bibitem [{\citenamefont {Albrecht}\ \emph {et~al.}(2016)\citenamefont
		{Albrecht}, \citenamefont {Higginbotham}, \citenamefont {Madsen},
		\citenamefont {Kuemmeth}, \citenamefont {Jespersen}, \citenamefont
		{Nyg{\aa}rd}, \citenamefont {Krogstrup},\ and\ \citenamefont
		{Marcus}}]{Albrecht2016Mar}%
	\BibitemOpen
	\bibfield  {author} {\bibinfo {author} {\bibfnamefont {S.~M.}\ \bibnamefont
			{Albrecht}}, \bibinfo {author} {\bibfnamefont {A.~P.}\ \bibnamefont
			{Higginbotham}}, \bibinfo {author} {\bibfnamefont {M.}~\bibnamefont
			{Madsen}}, \bibinfo {author} {\bibfnamefont {F.}~\bibnamefont {Kuemmeth}},
		\bibinfo {author} {\bibfnamefont {T.~S.}\ \bibnamefont {Jespersen}}, \bibinfo
		{author} {\bibfnamefont {J.}~\bibnamefont {Nyg{\aa}rd}}, \bibinfo {author}
		{\bibfnamefont {P.}~\bibnamefont {Krogstrup}},\ and\ \bibinfo {author}
		{\bibfnamefont {C.~M.}\ \bibnamefont {Marcus}},\ }\bibfield  {title}
	{\bibinfo {title} {{Exponential protection of zero modes in Majorana
				islands}},\ }\href {https://doi.org/10.1038/nature17162} {\bibfield
		{journal} {\bibinfo  {journal} {Nature}\ }\textbf {\bibinfo {volume} {531}},\
		\bibinfo {pages} {206} (\bibinfo {year} {2016})}\BibitemShut {NoStop}%
	\bibitem [{\citenamefont {Deng}\ \emph {et~al.}(2016)\citenamefont {Deng},
		\citenamefont {Vaitiek{\ifmmode\dot{e}\else\.{e}\fi}nas}, \citenamefont
		{Hansen}, \citenamefont {Danon}, \citenamefont {Leijnse}, \citenamefont
		{Flensberg}, \citenamefont {Nyg{\aa}rd}, \citenamefont {Krogstrup},\ and\
		\citenamefont {Marcus}}]{Deng2016Dec}%
	\BibitemOpen
	\bibfield  {author} {\bibinfo {author} {\bibfnamefont {M.~T.}\ \bibnamefont
			{Deng}}, \bibinfo {author} {\bibfnamefont {S.}~\bibnamefont
			{Vaitiek{\ifmmode\dot{e}\else\.{e}\fi}nas}}, \bibinfo {author} {\bibfnamefont
			{E.~B.}\ \bibnamefont {Hansen}}, \bibinfo {author} {\bibfnamefont
			{J.}~\bibnamefont {Danon}}, \bibinfo {author} {\bibfnamefont
			{M.}~\bibnamefont {Leijnse}}, \bibinfo {author} {\bibfnamefont
			{K.}~\bibnamefont {Flensberg}}, \bibinfo {author} {\bibfnamefont
			{J.}~\bibnamefont {Nyg{\aa}rd}}, \bibinfo {author} {\bibfnamefont
			{P.}~\bibnamefont {Krogstrup}},\ and\ \bibinfo {author} {\bibfnamefont
			{C.~M.}\ \bibnamefont {Marcus}},\ }\bibfield  {title} {\bibinfo {title}
		{{Majorana bound state in a coupled quantum-dot hybrid-nanowire system}},\
	}\href {https://doi.org/10.1126/science.aaf3961} {\bibfield  {journal}
		{\bibinfo  {journal} {Science}\ }\textbf {\bibinfo {volume} {354}},\ \bibinfo
		{pages} {1557} (\bibinfo {year} {2016})}\BibitemShut {NoStop}%
	\bibitem [{\citenamefont {Jeon}\ \emph {et~al.}(2017)\citenamefont {Jeon},
		\citenamefont {Xie}, \citenamefont {Li}, \citenamefont {Wang}, \citenamefont
		{Bernevig},\ and\ \citenamefont
		{Yazdani}}]{jeon_DistinguishingMajorana_2017}%
	\BibitemOpen
	\bibfield  {author} {\bibinfo {author} {\bibfnamefont {S.}~\bibnamefont
			{Jeon}}, \bibinfo {author} {\bibfnamefont {Y.}~\bibnamefont {Xie}}, \bibinfo
		{author} {\bibfnamefont {J.}~\bibnamefont {Li}}, \bibinfo {author}
		{\bibfnamefont {Z.}~\bibnamefont {Wang}}, \bibinfo {author} {\bibfnamefont
			{B.~A.}\ \bibnamefont {Bernevig}},\ and\ \bibinfo {author} {\bibfnamefont
			{A.}~\bibnamefont {Yazdani}},\ }\bibfield  {title} {\bibinfo {title}
		{Distinguishing a {{Majorana}} zero mode using spin-resolved measurements},\
	}\href {https://doi.org/10.1126/science.aan3670} {\bibfield  {journal}
		{\bibinfo  {journal} {Science}\ }\textbf {\bibinfo {volume} {358}},\ \bibinfo
		{pages} {772} (\bibinfo {year} {2017})}\BibitemShut {NoStop}%
	\bibitem [{\citenamefont {Nichele}\ \emph {et~al.}(2017)\citenamefont
		{Nichele}, \citenamefont {Drachmann}, \citenamefont {Whiticar}, \citenamefont
		{O'Farrell}, \citenamefont {Suominen}, \citenamefont {Fornieri},
		\citenamefont {Wang}, \citenamefont {Gardner}, \citenamefont {Thomas},
		\citenamefont {Hatke}, \citenamefont {Krogstrup}, \citenamefont {Manfra},
		\citenamefont {Flensberg},\ and\ \citenamefont
		{Marcus}}]{nichele_ScalingMajorana_2017}%
	\BibitemOpen
	\bibfield  {author} {\bibinfo {author} {\bibfnamefont {F.}~\bibnamefont
			{Nichele}}, \bibinfo {author} {\bibfnamefont {A.~C.~C.}\ \bibnamefont
			{Drachmann}}, \bibinfo {author} {\bibfnamefont {A.~M.}\ \bibnamefont
			{Whiticar}}, \bibinfo {author} {\bibfnamefont {E.~C.~T.}\ \bibnamefont
			{O'Farrell}}, \bibinfo {author} {\bibfnamefont {H.~J.}\ \bibnamefont
			{Suominen}}, \bibinfo {author} {\bibfnamefont {A.}~\bibnamefont {Fornieri}},
		\bibinfo {author} {\bibfnamefont {T.}~\bibnamefont {Wang}}, \bibinfo {author}
		{\bibfnamefont {G.~C.}\ \bibnamefont {Gardner}}, \bibinfo {author}
		{\bibfnamefont {C.}~\bibnamefont {Thomas}}, \bibinfo {author} {\bibfnamefont
			{A.~T.}\ \bibnamefont {Hatke}}, \bibinfo {author} {\bibfnamefont
			{P.}~\bibnamefont {Krogstrup}}, \bibinfo {author} {\bibfnamefont {M.~J.}\
			\bibnamefont {Manfra}}, \bibinfo {author} {\bibfnamefont {K.}~\bibnamefont
			{Flensberg}},\ and\ \bibinfo {author} {\bibfnamefont {C.~M.}\ \bibnamefont
			{Marcus}},\ }\bibfield  {title} {\bibinfo {title} {Scaling of {{Majorana
					Zero}}-{{Bias Conductance Peaks}}},\ }\href
	{https://doi.org/10.1103/PhysRevLett.119.136803} {\bibfield  {journal}
		{\bibinfo  {journal} {Physical Review Letters}\ }\textbf {\bibinfo {volume}
			{119}},\ \bibinfo {pages} {136803} (\bibinfo {year} {2017})},\ \Eprint
	{https://arxiv.org/abs/1706.07033} {arXiv:1706.07033} \BibitemShut {NoStop}%
	\bibitem [{\citenamefont {Deng}\ \emph {et~al.}(2018)\citenamefont {Deng},
		\citenamefont {Vaitiek{\ifmmode\dot{e}\else\.{e}\fi}nas}, \citenamefont
		{Prada}, \citenamefont {San-Jose}, \citenamefont {Nyg{\aa}rd}, \citenamefont
		{Krogstrup}, \citenamefont {Aguado},\ and\ \citenamefont
		{Marcus}}]{Deng2018Aug}%
	\BibitemOpen
	\bibfield  {author} {\bibinfo {author} {\bibfnamefont {M.-T.}\ \bibnamefont
			{Deng}}, \bibinfo {author} {\bibfnamefont {S.}~\bibnamefont
			{Vaitiek{\ifmmode\dot{e}\else\.{e}\fi}nas}}, \bibinfo {author} {\bibfnamefont
			{E.}~\bibnamefont {Prada}}, \bibinfo {author} {\bibfnamefont
			{P.}~\bibnamefont {San-Jose}}, \bibinfo {author} {\bibfnamefont
			{J.}~\bibnamefont {Nyg{\aa}rd}}, \bibinfo {author} {\bibfnamefont
			{P.}~\bibnamefont {Krogstrup}}, \bibinfo {author} {\bibfnamefont
			{R.}~\bibnamefont {Aguado}},\ and\ \bibinfo {author} {\bibfnamefont {C.~M.}\
			\bibnamefont {Marcus}},\ }\bibfield  {title} {\bibinfo {title} {{Nonlocality
				of Majorana modes in hybrid nanowires}},\ }\href
	{https://doi.org/10.1103/PhysRevB.98.085125} {\bibfield  {journal} {\bibinfo
			{journal} {Phys. Rev. B}\ }\textbf {\bibinfo {volume} {98}},\ \bibinfo
		{pages} {085125} (\bibinfo {year} {2018})}\BibitemShut {NoStop}%
	\bibitem [{\citenamefont {Lutchyn}\ \emph {et~al.}(2018)\citenamefont
		{Lutchyn}, \citenamefont {Bakkers}, \citenamefont {Kouwenhoven},
		\citenamefont {Krogstrup}, \citenamefont {Marcus},\ and\ \citenamefont
		{Oreg}}]{Lutchyn2018May}%
	\BibitemOpen
	\bibfield  {author} {\bibinfo {author} {\bibfnamefont {R.~M.}\ \bibnamefont
			{Lutchyn}}, \bibinfo {author} {\bibfnamefont {E.~P. A.~M.}\ \bibnamefont
			{Bakkers}}, \bibinfo {author} {\bibfnamefont {L.~P.}\ \bibnamefont
			{Kouwenhoven}}, \bibinfo {author} {\bibfnamefont {P.}~\bibnamefont
			{Krogstrup}}, \bibinfo {author} {\bibfnamefont {C.~M.}\ \bibnamefont
			{Marcus}},\ and\ \bibinfo {author} {\bibfnamefont {Y.}~\bibnamefont {Oreg}},\
	}\bibfield  {title} {\bibinfo {title} {{Majorana zero modes in
				superconductor{\textendash}semiconductor heterostructures}},\ }\href
	{https://doi.org/10.1038/s41578-018-0003-1} {\bibfield  {journal} {\bibinfo
			{journal} {Nat. Rev. Mater.}\ }\textbf {\bibinfo {volume} {3}},\ \bibinfo
		{pages} {52} (\bibinfo {year} {2018})}\BibitemShut {NoStop}%
	\bibitem [{\citenamefont {G{\ifmmode\ddot{u}\else\"{u}\fi}l}\ \emph
		{et~al.}(2018)\citenamefont {G{\ifmmode\ddot{u}\else\"{u}\fi}l},
		\citenamefont {Zhang}, \citenamefont {Bommer}, \citenamefont {de~Moor},
		\citenamefont {Car}, \citenamefont {Plissard}, \citenamefont {Bakkers},
		\citenamefont {Geresdi}, \citenamefont {Watanabe}, \citenamefont
		{Taniguchi},\ and\ \citenamefont {Kouwenhoven}}]{Gul2018Jan}%
	\BibitemOpen
	\bibfield  {author} {\bibinfo {author} {\bibfnamefont
			{{\ifmmode\ddot{O}\else\"{O}\fi}.}~\bibnamefont
			{G{\ifmmode\ddot{u}\else\"{u}\fi}l}}, \bibinfo {author} {\bibfnamefont
			{H.}~\bibnamefont {Zhang}}, \bibinfo {author} {\bibfnamefont {J.~D.~S.}\
			\bibnamefont {Bommer}}, \bibinfo {author} {\bibfnamefont {M.~W.~A.}\
			\bibnamefont {de~Moor}}, \bibinfo {author} {\bibfnamefont {D.}~\bibnamefont
			{Car}}, \bibinfo {author} {\bibfnamefont {S.~R.}\ \bibnamefont {Plissard}},
		\bibinfo {author} {\bibfnamefont {E.~P. A.~M.}\ \bibnamefont {Bakkers}},
		\bibinfo {author} {\bibfnamefont {A.}~\bibnamefont {Geresdi}}, \bibinfo
		{author} {\bibfnamefont {K.}~\bibnamefont {Watanabe}}, \bibinfo {author}
		{\bibfnamefont {T.}~\bibnamefont {Taniguchi}},\ and\ \bibinfo {author}
		{\bibfnamefont {L.~P.}\ \bibnamefont {Kouwenhoven}},\ }\bibfield  {title}
	{\bibinfo {title} {{Ballistic Majorana nanowire devices}},\ }\href
	{https://doi.org/10.1038/s41565-017-0032-8} {\bibfield  {journal} {\bibinfo
			{journal} {Nat. Nanotechnol.}\ }\textbf {\bibinfo {volume} {13}},\ \bibinfo
		{pages} {192} (\bibinfo {year} {2018})}\BibitemShut {NoStop}%
	\bibitem [{\citenamefont {Zhang}\ \emph {et~al.}(2019)\citenamefont {Zhang},
		\citenamefont {Liu}, \citenamefont {Wimmer},\ and\ \citenamefont
		{Kouwenhoven}}]{zhang_NextSteps_2019}%
	\BibitemOpen
	\bibfield  {author} {\bibinfo {author} {\bibfnamefont {H.}~\bibnamefont
			{Zhang}}, \bibinfo {author} {\bibfnamefont {D.~E.}\ \bibnamefont {Liu}},
		\bibinfo {author} {\bibfnamefont {M.}~\bibnamefont {Wimmer}},\ and\ \bibinfo
		{author} {\bibfnamefont {L.~P.}\ \bibnamefont {Kouwenhoven}},\ }\bibfield
	{title} {\bibinfo {title} {Next steps of quantum transport in {{Majorana}}
			nanowire devices},\ }\href {https://doi.org/10.1038/s41467-019-13133-1}
	{\bibfield  {journal} {\bibinfo  {journal} {Nature Communications}\ }\textbf
		{\bibinfo {volume} {10}},\ \bibinfo {pages} {5128} (\bibinfo {year}
		{2019})}\BibitemShut {NoStop}%
	\bibitem [{\citenamefont {Kitaev}(2003)}]{kitaev_fault-tolerant_2003}%
	\BibitemOpen
	\bibfield  {author} {\bibinfo {author} {\bibfnamefont {A.~Y.}\ \bibnamefont
			{Kitaev}},\ }\bibfield  {title} {\bibinfo {title} {Fault-tolerant quantum
			computation by anyons},\ }\href
	{https://doi.org/10.1016/S0003-4916(02)00018-0} {\bibfield  {journal}
		{\bibinfo  {journal} {Ann. Phys.}\ }\textbf {\bibinfo {volume} {303}},\
		\bibinfo {pages} {2} (\bibinfo {year} {2003})},\ \bibinfo {note} {publisher:
		Academic Press}\BibitemShut {NoStop}%
	\bibitem [{\citenamefont {Nayak}\ \emph {et~al.}(2008)\citenamefont {Nayak},
		\citenamefont {Simon}, \citenamefont {Stern}, \citenamefont {Freedman},\ and\
		\citenamefont {Das~Sarma}}]{Nayak2008Sep}%
	\BibitemOpen
	\bibfield  {author} {\bibinfo {author} {\bibfnamefont {C.}~\bibnamefont
			{Nayak}}, \bibinfo {author} {\bibfnamefont {S.~H.}\ \bibnamefont {Simon}},
		\bibinfo {author} {\bibfnamefont {A.}~\bibnamefont {Stern}}, \bibinfo
		{author} {\bibfnamefont {M.}~\bibnamefont {Freedman}},\ and\ \bibinfo
		{author} {\bibfnamefont {S.}~\bibnamefont {Das~Sarma}},\ }\bibfield  {title}
	{\bibinfo {title} {{Non-Abelian anyons and topological quantum
				computation}},\ }\href {https://doi.org/10.1103/RevModPhys.80.1083}
	{\bibfield  {journal} {\bibinfo  {journal} {Rev. Mod. Phys.}\ }\textbf
		{\bibinfo {volume} {80}},\ \bibinfo {pages} {1083} (\bibinfo {year}
		{2008})}\BibitemShut {NoStop}%
	\bibitem [{\citenamefont {Alicea}\ \emph {et~al.}(2011)\citenamefont {Alicea},
		\citenamefont {Oreg}, \citenamefont {Refael}, \citenamefont {von Oppen},\
		and\ \citenamefont {Fisher}}]{alicea_NonAbelianStatistics_2011}%
	\BibitemOpen
	\bibfield  {author} {\bibinfo {author} {\bibfnamefont {J.}~\bibnamefont
			{Alicea}}, \bibinfo {author} {\bibfnamefont {Y.}~\bibnamefont {Oreg}},
		\bibinfo {author} {\bibfnamefont {G.}~\bibnamefont {Refael}}, \bibinfo
		{author} {\bibfnamefont {F.}~\bibnamefont {von Oppen}},\ and\ \bibinfo
		{author} {\bibfnamefont {M.~P.~A.}\ \bibnamefont {Fisher}},\ }\bibfield
	{title} {\bibinfo {title} {Non-{{Abelian}} statistics and topological quantum
			information processing in {{1D}} wire networks},\ }\href
	{https://doi.org/10.1038/nphys1915} {\bibfield  {journal} {\bibinfo
			{journal} {Nature Physics}\ }\textbf {\bibinfo {volume} {7}},\ \bibinfo
		{pages} {412} (\bibinfo {year} {2011})}\BibitemShut {NoStop}%
	\bibitem [{\citenamefont {Liu}\ and\ \citenamefont
		{Baranger}(2011)}]{Liu2011Nov}%
	\BibitemOpen
	\bibfield  {author} {\bibinfo {author} {\bibfnamefont {D.~E.}\ \bibnamefont
			{Liu}}\ and\ \bibinfo {author} {\bibfnamefont {H.~U.}\ \bibnamefont
			{Baranger}},\ }\bibfield  {title} {\bibinfo {title} {{Detecting a
				Majorana-fermion zero mode using a quantum dot}},\ }\href
	{https://doi.org/10.1103/PhysRevB.84.201308} {\bibfield  {journal} {\bibinfo
			{journal} {Phys. Rev. B}\ }\textbf {\bibinfo {volume} {84}},\ \bibinfo
		{pages} {201308(R)} (\bibinfo {year} {2011})}\BibitemShut {NoStop}%
	\bibitem [{\citenamefont {Leijnse}\ and\ \citenamefont
		{Flensberg}(2011)}]{Leijnse2011Oct}%
	\BibitemOpen
	\bibfield  {author} {\bibinfo {author} {\bibfnamefont {M.}~\bibnamefont
			{Leijnse}}\ and\ \bibinfo {author} {\bibfnamefont {K.}~\bibnamefont
			{Flensberg}},\ }\bibfield  {title} {\bibinfo {title} {{Scheme to measure
				Majorana fermion lifetimes using a quantum dot}},\ }\href
	{https://doi.org/10.1103/PhysRevB.84.140501} {\bibfield  {journal} {\bibinfo
			{journal} {Phys. Rev. B}\ }\textbf {\bibinfo {volume} {84}},\ \bibinfo
		{pages} {140501(R)} (\bibinfo {year} {2011})}\BibitemShut {NoStop}%
	\bibitem [{\citenamefont {Cao}\ \emph {et~al.}(2012)\citenamefont {Cao},
		\citenamefont {Wang}, \citenamefont {Xiong}, \citenamefont {Gong},\ and\
		\citenamefont {Li}}]{Cao2012Sep}%
	\BibitemOpen
	\bibfield  {author} {\bibinfo {author} {\bibfnamefont {Y.}~\bibnamefont
			{Cao}}, \bibinfo {author} {\bibfnamefont {P.}~\bibnamefont {Wang}}, \bibinfo
		{author} {\bibfnamefont {G.}~\bibnamefont {Xiong}}, \bibinfo {author}
		{\bibfnamefont {M.}~\bibnamefont {Gong}},\ and\ \bibinfo {author}
		{\bibfnamefont {X.-Q.}\ \bibnamefont {Li}},\ }\bibfield  {title} {\bibinfo
		{title} {{Probing the existence and dynamics of Majorana fermion via
				transport through a quantum dot}},\ }\href
	{https://doi.org/10.1103/PhysRevB.86.115311} {\bibfield  {journal} {\bibinfo
			{journal} {Phys. Rev. B}\ }\textbf {\bibinfo {volume} {86}},\ \bibinfo
		{pages} {115311} (\bibinfo {year} {2012})}\BibitemShut {NoStop}%
	\bibitem [{\citenamefont {Gong}\ \emph {et~al.}(2014)\citenamefont {Gong},
		\citenamefont {Zhang}, \citenamefont {Li}, \citenamefont {Yi},\ and\
		\citenamefont {Zheng}}]{Gong2014Jun}%
	\BibitemOpen
	\bibfield  {author} {\bibinfo {author} {\bibfnamefont {W.-J.}\ \bibnamefont
			{Gong}}, \bibinfo {author} {\bibfnamefont {S.-F.}\ \bibnamefont {Zhang}},
		\bibinfo {author} {\bibfnamefont {Z.-C.}\ \bibnamefont {Li}}, \bibinfo
		{author} {\bibfnamefont {G.}~\bibnamefont {Yi}},\ and\ \bibinfo {author}
		{\bibfnamefont {Y.-S.}\ \bibnamefont {Zheng}},\ }\bibfield  {title} {\bibinfo
		{title} {{Detection of a Majorana fermion zero mode by a T-shaped quantum-dot
				structure}},\ }\href {https://doi.org/10.1103/PhysRevB.89.245413} {\bibfield
		{journal} {\bibinfo  {journal} {Phys. Rev. B}\ }\textbf {\bibinfo {volume}
			{89}},\ \bibinfo {pages} {245413} (\bibinfo {year} {2014})}\BibitemShut
	{NoStop}%
	\bibitem [{\citenamefont {Cheng}\ \emph {et~al.}(2014)\citenamefont {Cheng},
		\citenamefont {Becker}, \citenamefont {Bauer},\ and\ \citenamefont
		{Lutchyn}}]{Cheng2014Sep}%
	\BibitemOpen
	\bibfield  {author} {\bibinfo {author} {\bibfnamefont {M.}~\bibnamefont
			{Cheng}}, \bibinfo {author} {\bibfnamefont {M.}~\bibnamefont {Becker}},
		\bibinfo {author} {\bibfnamefont {B.}~\bibnamefont {Bauer}},\ and\ \bibinfo
		{author} {\bibfnamefont {R.~M.}\ \bibnamefont {Lutchyn}},\ }\bibfield
	{title} {\bibinfo {title} {{Interplay between Kondo and Majorana Interactions
				in Quantum Dots}},\ }\href {https://doi.org/10.1103/PhysRevX.4.031051}
	{\bibfield  {journal} {\bibinfo  {journal} {Phys. Rev. X}\ }\textbf {\bibinfo
			{volume} {4}},\ \bibinfo {pages} {031051} (\bibinfo {year}
		{2014})}\BibitemShut {NoStop}%
	\bibitem [{\citenamefont {Liu}\ \emph {et~al.}(2015)\citenamefont {Liu},
		\citenamefont {Cheng},\ and\ \citenamefont {Lutchyn}}]{Liu2015Feb}%
	\BibitemOpen
	\bibfield  {author} {\bibinfo {author} {\bibfnamefont {D.~E.}\ \bibnamefont
			{Liu}}, \bibinfo {author} {\bibfnamefont {M.}~\bibnamefont {Cheng}},\ and\
		\bibinfo {author} {\bibfnamefont {R.~M.}\ \bibnamefont {Lutchyn}},\
	}\bibfield  {title} {\bibinfo {title} {{Probing Majorana physics in
				quantum-dot shot-noise experiments}},\ }\href
	{https://doi.org/10.1103/PhysRevB.91.081405} {\bibfield  {journal} {\bibinfo
			{journal} {Phys. Rev. B}\ }\textbf {\bibinfo {volume} {91}},\ \bibinfo
		{pages} {081405(R)} (\bibinfo {year} {2015})}\BibitemShut {NoStop}%
	\bibitem [{\citenamefont {Schuray}\ \emph {et~al.}(2017)\citenamefont
		{Schuray}, \citenamefont {Weithofer},\ and\ \citenamefont
		{Recher}}]{schuray_FanoResonances_2017}%
	\BibitemOpen
	\bibfield  {author} {\bibinfo {author} {\bibfnamefont {A.}~\bibnamefont
			{Schuray}}, \bibinfo {author} {\bibfnamefont {L.}~\bibnamefont {Weithofer}},\
		and\ \bibinfo {author} {\bibfnamefont {P.}~\bibnamefont {Recher}},\
	}\bibfield  {title} {\bibinfo {title} {Fano {{Resonances}} in {{Majorana
					Bound States}} - {{Quantum Dot Hybrid Systems}}},\ }\href
	{https://doi.org/10.1103/PhysRevB.96.085417} {\bibfield  {journal} {\bibinfo
			{journal} {Physical Review B}\ }\textbf {\bibinfo {volume} {96}},\ \bibinfo
		{pages} {085417} (\bibinfo {year} {2017})},\ \Eprint
	{https://arxiv.org/abs/1702.03909} {arXiv:1702.03909} \BibitemShut {NoStop}%
	\bibitem [{\citenamefont {Prada}\ \emph {et~al.}(2017)\citenamefont {Prada},
		\citenamefont {Aguado},\ and\ \citenamefont
		{{San-Jose}}}]{prada_MeasuringMajorana_2017}%
	\BibitemOpen
	\bibfield  {author} {\bibinfo {author} {\bibfnamefont {E.}~\bibnamefont
			{Prada}}, \bibinfo {author} {\bibfnamefont {R.}~\bibnamefont {Aguado}},\ and\
		\bibinfo {author} {\bibfnamefont {P.}~\bibnamefont {{San-Jose}}},\ }\bibfield
	{title} {\bibinfo {title} {Measuring {{Majorana}} nonlocality and spin
			structure with a quantum dot},\ }\href
	{https://doi.org/10.1103/PhysRevB.96.085418} {\bibfield  {journal} {\bibinfo
			{journal} {Physical Review B}\ }\textbf {\bibinfo {volume} {96}},\ \bibinfo
		{pages} {085418} (\bibinfo {year} {2017})}\BibitemShut {NoStop}%
	\bibitem [{\citenamefont {Hoffman}\ \emph {et~al.}(2017)\citenamefont
		{Hoffman}, \citenamefont {Chevallier}, \citenamefont {Loss},\ and\
		\citenamefont {Klinovaja}}]{hoffman_SpindependentCoupling_2017}%
	\BibitemOpen
	\bibfield  {author} {\bibinfo {author} {\bibfnamefont {S.}~\bibnamefont
			{Hoffman}}, \bibinfo {author} {\bibfnamefont {D.}~\bibnamefont {Chevallier}},
		\bibinfo {author} {\bibfnamefont {D.}~\bibnamefont {Loss}},\ and\ \bibinfo
		{author} {\bibfnamefont {J.}~\bibnamefont {Klinovaja}},\ }\bibfield  {title}
	{\bibinfo {title} {Spin-dependent coupling between quantum dots and
			topological quantum wires},\ }\href
	{https://doi.org/10.1103/PhysRevB.96.045440} {\bibfield  {journal} {\bibinfo
			{journal} {Phys. Rev. B}\ }\textbf {\bibinfo {volume} {96}},\ \bibinfo
		{pages} {045440} (\bibinfo {year} {2017})}\BibitemShut {NoStop}%
	\bibitem [{\citenamefont {G{\'o}rski}\ \emph {et~al.}(2018)\citenamefont
		{G{\'o}rski}, \citenamefont {Bara{\'n}ski}, \citenamefont {Weymann},\ and\
		\citenamefont {Doma{\'n}ski}}]{gorski_InterplayCorrelations_2018}%
	\BibitemOpen
	\bibfield  {author} {\bibinfo {author} {\bibfnamefont {G.}~\bibnamefont
			{G{\'o}rski}}, \bibinfo {author} {\bibfnamefont {J.}~\bibnamefont
			{Bara{\'n}ski}}, \bibinfo {author} {\bibfnamefont {I.}~\bibnamefont
			{Weymann}},\ and\ \bibinfo {author} {\bibfnamefont {T.}~\bibnamefont
			{Doma{\'n}ski}},\ }\bibfield  {title} {\bibinfo {title} {Interplay between
			correlations and {{Majorana}} mode in proximitized quantum dot},\ }\href
	{https://doi.org/10.1038/s41598-018-33529-1} {\bibfield  {journal} {\bibinfo
			{journal} {Scientific Reports}\ }\textbf {\bibinfo {volume} {8}},\ \bibinfo
		{pages} {15717} (\bibinfo {year} {2018})}\BibitemShut {NoStop}%
	\bibitem [{\citenamefont {Sanches}\ \emph {et~al.}(2020)\citenamefont
		{Sanches}, \citenamefont {Ricco}, \citenamefont {Mizobata}, \citenamefont
		{Marques}, \citenamefont {de~Souza}, \citenamefont {Shelykh},\ and\
		\citenamefont {Seridonio}}]{sanches_MajoranaMolecules_2020}%
	\BibitemOpen
	\bibfield  {author} {\bibinfo {author} {\bibfnamefont {J.~E.}\ \bibnamefont
			{Sanches}}, \bibinfo {author} {\bibfnamefont {L.~S.}\ \bibnamefont {Ricco}},
		\bibinfo {author} {\bibfnamefont {W.~N.}\ \bibnamefont {Mizobata}}, \bibinfo
		{author} {\bibfnamefont {Y.}~\bibnamefont {Marques}}, \bibinfo {author}
		{\bibfnamefont {M.}~\bibnamefont {de~Souza}}, \bibinfo {author}
		{\bibfnamefont {I.~A.}\ \bibnamefont {Shelykh}},\ and\ \bibinfo {author}
		{\bibfnamefont {A.~C.}\ \bibnamefont {Seridonio}},\ }\bibfield  {title}
	{\bibinfo {title} {Majorana molecules and their spectral fingerprints},\
	}\href {https://doi.org/10.1103/PhysRevB.102.075128} {\bibfield  {journal}
		{\bibinfo  {journal} {Physical Review B}\ }\textbf {\bibinfo {volume}
			{102}},\ \bibinfo {pages} {075128} (\bibinfo {year} {2020})},\ \Eprint
	{https://arxiv.org/abs/2004.00374} {arXiv:2004.00374} \BibitemShut {NoStop}%
	\bibitem [{\citenamefont {Wrze{\'s}niewski}\ and\ \citenamefont
		{Weymann}(2021)}]{wrzesniewski_MagnetizationDynamics_2021}%
	\BibitemOpen
	\bibfield  {author} {\bibinfo {author} {\bibfnamefont {K.}~\bibnamefont
			{Wrze{\'s}niewski}}\ and\ \bibinfo {author} {\bibfnamefont {I.}~\bibnamefont
			{Weymann}},\ }\bibfield  {title} {\bibinfo {title} {Magnetization dynamics in
			a {{Majorana}}-wire--quantum-dot setup},\ }\href
	{https://doi.org/10.1103/PhysRevB.103.125413} {\bibfield  {journal} {\bibinfo
			{journal} {Physical Review B}\ }\textbf {\bibinfo {volume} {103}},\ \bibinfo
		{pages} {125413} (\bibinfo {year} {2021})}\BibitemShut {NoStop}%
	\bibitem [{\citenamefont {Vernek}\ \emph {et~al.}(2014)\citenamefont {Vernek},
		\citenamefont {Penteado}, \citenamefont {Seridonio},\ and\ \citenamefont
		{Egues}}]{Vernek2014Apr}%
	\BibitemOpen
	\bibfield  {author} {\bibinfo {author} {\bibfnamefont {E.}~\bibnamefont
			{Vernek}}, \bibinfo {author} {\bibfnamefont {P.~H.}\ \bibnamefont
			{Penteado}}, \bibinfo {author} {\bibfnamefont {A.~C.}\ \bibnamefont
			{Seridonio}},\ and\ \bibinfo {author} {\bibfnamefont {J.~C.}\ \bibnamefont
			{Egues}},\ }\bibfield  {title} {\bibinfo {title} {{Subtle leakage of a
				Majorana mode into a quantum dot}},\ }\href
	{https://doi.org/10.1103/PhysRevB.89.165314} {\bibfield  {journal} {\bibinfo
			{journal} {Phys. Rev. B}\ }\textbf {\bibinfo {volume} {89}},\ \bibinfo
		{pages} {165314} (\bibinfo {year} {2014})}\BibitemShut {NoStop}%
	\bibitem [{\citenamefont {Ruiz-Tijerina}\ \emph {et~al.}(2015)\citenamefont
		{Ruiz-Tijerina}, \citenamefont {Vernek}, \citenamefont {Dias~da Silva},\ and\
		\citenamefont {Egues}}]{Ruiz-Tijerina2015Mar}%
	\BibitemOpen
	\bibfield  {author} {\bibinfo {author} {\bibfnamefont {D.~A.}\ \bibnamefont
			{Ruiz-Tijerina}}, \bibinfo {author} {\bibfnamefont {E.}~\bibnamefont
			{Vernek}}, \bibinfo {author} {\bibfnamefont {L.~G. G.~V.}\ \bibnamefont
			{Dias~da Silva}},\ and\ \bibinfo {author} {\bibfnamefont {J.~C.}\
			\bibnamefont {Egues}},\ }\bibfield  {title} {\bibinfo {title} {{Interaction
				effects on a Majorana zero mode leaking into a quantum dot}},\ }\href
	{https://doi.org/10.1103/PhysRevB.91.115435} {\bibfield  {journal} {\bibinfo
			{journal} {Phys. Rev. B}\ }\textbf {\bibinfo {volume} {91}},\ \bibinfo
		{pages} {115435} (\bibinfo {year} {2015})}\BibitemShut {NoStop}%
	\bibitem [{\citenamefont {Liu}\ \emph {et~al.}(2017)\citenamefont {Liu},
		\citenamefont {Sau}, \citenamefont {Stanescu},\ and\ \citenamefont
		{Das~Sarma}}]{liu_AndreevBound_2017}%
	\BibitemOpen
	\bibfield  {author} {\bibinfo {author} {\bibfnamefont {C.-X.}\ \bibnamefont
			{Liu}}, \bibinfo {author} {\bibfnamefont {J.~D.}\ \bibnamefont {Sau}},
		\bibinfo {author} {\bibfnamefont {T.~D.}\ \bibnamefont {Stanescu}},\ and\
		\bibinfo {author} {\bibfnamefont {S.}~\bibnamefont {Das~Sarma}},\ }\bibfield
	{title} {\bibinfo {title} {Andreev bound states versus {{Majorana}} bound
			states in quantum dot-nanowire-superconductor hybrid structures: {{Trivial}}
			versus topological zero-bias conductance peaks},\ }\href
	{https://doi.org/10.1103/PhysRevB.96.075161} {\bibfield  {journal} {\bibinfo
			{journal} {Physical Review B}\ }\textbf {\bibinfo {volume} {96}},\ \bibinfo
		{pages} {075161} (\bibinfo {year} {2017})}\BibitemShut {NoStop}%
	\bibitem [{\citenamefont {Zienkiewicz}\ \emph {et~al.}(2019)\citenamefont
		{Zienkiewicz}, \citenamefont {Bara{\'n}ski}, \citenamefont {G{\'o}rski},\
		and\ \citenamefont {Doma{\'n}ski}}]{zienkiewicz_LeakageMajorana_2019}%
	\BibitemOpen
	\bibfield  {author} {\bibinfo {author} {\bibfnamefont {T.}~\bibnamefont
			{Zienkiewicz}}, \bibinfo {author} {\bibfnamefont {J.}~\bibnamefont
			{Bara{\'n}ski}}, \bibinfo {author} {\bibfnamefont {G.}~\bibnamefont
			{G{\'o}rski}},\ and\ \bibinfo {author} {\bibfnamefont {T.}~\bibnamefont
			{Doma{\'n}ski}},\ }\bibfield  {title} {\bibinfo {title} {Leakage of
			{{Majorana}} mode into correlated quantum dot nearby its singlet-doublet
			crossover},\ }\href {https://doi.org/10.1088/1361-648x/ab46d9} {\bibfield
		{journal} {\bibinfo  {journal} {J. Phys.: Condens. Matter}\ }\textbf
		{\bibinfo {volume} {32}},\ \bibinfo {pages} {025302} (\bibinfo {year}
		{2019})}\BibitemShut {NoStop}%
	\bibitem [{\citenamefont {Lee}\ \emph {et~al.}(2013)\citenamefont {Lee},
		\citenamefont {Lim},\ and\ \citenamefont
		{L{\ifmmode\acute{o}\else\'{o}\fi}pez}}]{Lee2013Jun}%
	\BibitemOpen
	\bibfield  {author} {\bibinfo {author} {\bibfnamefont {M.}~\bibnamefont
			{Lee}}, \bibinfo {author} {\bibfnamefont {J.~S.}\ \bibnamefont {Lim}},\ and\
		\bibinfo {author} {\bibfnamefont {R.}~\bibnamefont
			{L{\ifmmode\acute{o}\else\'{o}\fi}pez}},\ }\bibfield  {title} {\bibinfo
		{title} {{Kondo effect in a quantum dot side-coupled to a topological
				superconductor}},\ }\href {https://doi.org/10.1103/PhysRevB.87.241402}
	{\bibfield  {journal} {\bibinfo  {journal} {Phys. Rev. B}\ }\textbf {\bibinfo
			{volume} {87}},\ \bibinfo {pages} {241402(R)} (\bibinfo {year}
		{2013})}\BibitemShut {NoStop}%
	\bibitem [{\citenamefont {Golub}\ \emph {et~al.}(2011)\citenamefont {Golub},
		\citenamefont {Kuzmenko},\ and\ \citenamefont {Avishai}}]{Golub2011Oct}%
	\BibitemOpen
	\bibfield  {author} {\bibinfo {author} {\bibfnamefont {A.}~\bibnamefont
			{Golub}}, \bibinfo {author} {\bibfnamefont {I.}~\bibnamefont {Kuzmenko}},\
		and\ \bibinfo {author} {\bibfnamefont {Y.}~\bibnamefont {Avishai}},\
	}\bibfield  {title} {\bibinfo {title} {{Kondo Correlations and Majorana Bound
				States in a Metal to Quantum-Dot to Topological-Superconductor Junction}},\
	}\href {https://doi.org/10.1103/PhysRevLett.107.176802} {\bibfield  {journal}
		{\bibinfo  {journal} {Phys. Rev. Lett.}\ }\textbf {\bibinfo {volume} {107}},\
		\bibinfo {pages} {176802} (\bibinfo {year} {2011})}\BibitemShut {NoStop}%
	\bibitem [{\citenamefont {Eriksson}\ \emph {et~al.}(2014)\citenamefont
		{Eriksson}, \citenamefont {Nava}, \citenamefont {Mora},\ and\ \citenamefont
		{Egger}}]{eriksson_TunnelingSpectroscopy_2014}%
	\BibitemOpen
	\bibfield  {author} {\bibinfo {author} {\bibfnamefont {E.}~\bibnamefont
			{Eriksson}}, \bibinfo {author} {\bibfnamefont {A.}~\bibnamefont {Nava}},
		\bibinfo {author} {\bibfnamefont {C.}~\bibnamefont {Mora}},\ and\ \bibinfo
		{author} {\bibfnamefont {R.}~\bibnamefont {Egger}},\ }\bibfield  {title}
	{\bibinfo {title} {Tunneling spectroscopy of {{Majorana}}-{{Kondo}}
			devices},\ }\href {https://doi.org/10.1103/PhysRevB.90.245417} {\bibfield
		{journal} {\bibinfo  {journal} {Physical Review B}\ }\textbf {\bibinfo
			{volume} {90}},\ \bibinfo {pages} {245417} (\bibinfo {year}
		{2014})}\BibitemShut {NoStop}%
	\bibitem [{\citenamefont {Weymann}\ and\ \citenamefont
		{W{\ifmmode\acute{o}\else\'{o}\fi}jcik}(2017)}]{Weymann2017Apr}%
	\BibitemOpen
	\bibfield  {author} {\bibinfo {author} {\bibfnamefont {I.}~\bibnamefont
			{Weymann}}\ and\ \bibinfo {author} {\bibfnamefont {K.~P.}\ \bibnamefont
			{W{\ifmmode\acute{o}\else\'{o}\fi}jcik}},\ }\bibfield  {title} {\bibinfo
		{title} {{Transport properties of a hybrid Majorana wire-quantum dot system
				with ferromagnetic contacts}},\ }\href
	{https://doi.org/10.1103/PhysRevB.95.155427} {\bibfield  {journal} {\bibinfo
			{journal} {Phys. Rev. B}\ }\textbf {\bibinfo {volume} {95}},\ \bibinfo
		{pages} {155427} (\bibinfo {year} {2017})}\BibitemShut {NoStop}%
	\bibitem [{\citenamefont {Weymann}(2017)}]{Weymann2017Jan}%
	\BibitemOpen
	\bibfield  {author} {\bibinfo {author} {\bibfnamefont {I.}~\bibnamefont
			{Weymann}},\ }\bibfield  {title} {\bibinfo {title} {{Spin Seebeck effect in
				quantum dot side-coupled to topological superconductor}},\ }\href
	{https://doi.org/10.1088/1361-648x/aa5526} {\bibfield  {journal} {\bibinfo
			{journal} {J. Phys.: Condens. Matter}\ }\textbf {\bibinfo {volume} {29}},\
		\bibinfo {pages} {095301} (\bibinfo {year} {2017})}\BibitemShut {NoStop}%
	\bibitem [{\citenamefont {Silva}\ \emph
		{et~al.}(2020{\natexlab{a}})\citenamefont {Silva}, \citenamefont {da~Silva},\
		and\ \citenamefont {Vernek}}]{Vernek2019}%
	\BibitemOpen
	\bibfield  {author} {\bibinfo {author} {\bibfnamefont {J.~F.}\ \bibnamefont
			{Silva}}, \bibinfo {author} {\bibfnamefont {L.~G. G. V.~D.}\ \bibnamefont
			{da~Silva}},\ and\ \bibinfo {author} {\bibfnamefont {E.}~\bibnamefont
			{Vernek}},\ }\bibfield  {title} {\bibinfo {title} {{Robustness of the Kondo
				effect in a quantum dot coupled to Majorana zero modes}},\ }\href
	{https://doi.org/10.1103/PhysRevB.101.075428} {\bibfield  {journal} {\bibinfo
			{journal} {Phys. Rev. B}\ }\textbf {\bibinfo {volume} {101}},\ \bibinfo
		{pages} {075428} (\bibinfo {year} {2020}{\natexlab{a}})}\BibitemShut
	{NoStop}%
	\bibitem [{\citenamefont {Weymann}\ \emph {et~al.}()\citenamefont {Weymann},
		\citenamefont {Wójcik},\ and\ \citenamefont
		{Majek}}]{weymann_majorana-kondo_2020}%
	\BibitemOpen
	\bibfield  {author} {\bibinfo {author} {\bibfnamefont {I.}~\bibnamefont
			{Weymann}}, \bibinfo {author} {\bibfnamefont {K.~P.}\ \bibnamefont
			{Wójcik}},\ and\ \bibinfo {author} {\bibfnamefont {P.}~\bibnamefont
			{Majek}},\ }\bibfield  {title} {\bibinfo {title} {Majorana-kondo interplay in
			t-shaped double quantum dots},\ }\href
	{https://doi.org/10.1103/PhysRevB.101.235404} {\bibfield  {journal} {\bibinfo
			{journal} {Phys. Rev. B}\ }\textbf {\bibinfo {volume} {101}},\ \bibinfo
		{pages} {235404}}\BibitemShut {NoStop}%
	\bibitem [{\citenamefont {Silva}\ \emph
		{et~al.}(2020{\natexlab{b}})\citenamefont {Silva}, \citenamefont {da~Silva},\
		and\ \citenamefont {Vernek}}]{silva_RobustnessKondo_2020}%
	\BibitemOpen
	\bibfield  {author} {\bibinfo {author} {\bibfnamefont {J.~F.}\ \bibnamefont
			{Silva}}, \bibinfo {author} {\bibfnamefont {L.~G. G. V.~D.}\ \bibnamefont
			{da~Silva}},\ and\ \bibinfo {author} {\bibfnamefont {E.}~\bibnamefont
			{Vernek}},\ }\bibfield  {title} {\bibinfo {title} {Robustness of the
			{{Kondo}} effect in a quantum dot coupled to {{Majorana}} zero modes},\
	}\href {https://doi.org/10.1103/PhysRevB.101.075428} {\bibfield  {journal}
		{\bibinfo  {journal} {Physical Review B}\ }\textbf {\bibinfo {volume}
			{101}},\ \bibinfo {pages} {075428} (\bibinfo {year}
		{2020}{\natexlab{b}})}\BibitemShut {NoStop}%
	\bibitem [{\citenamefont {Hewson}(1993)}]{hewson_1993}%
	\BibitemOpen
	\bibfield  {author} {\bibinfo {author} {\bibfnamefont {A.~C.}\ \bibnamefont
			{Hewson}},\ }\href {https://doi.org/10.1017/CBO9780511470752} {\emph
		{\bibinfo {title} {{The Kondo Problem to Heavy Fermions}}}},\ Cambridge
	Studies in Magnetism\ (\bibinfo  {publisher} {Cambridge University Press},\
	\bibinfo {year} {1993})\BibitemShut {NoStop}%
	\bibitem [{\citenamefont {Vojta}\ \emph {et~al.}(2002)\citenamefont {Vojta},
		\citenamefont {Bulla},\ and\ \citenamefont {Hofstetter}}]{vojtaPRB02}%
	\BibitemOpen
	\bibfield  {author} {\bibinfo {author} {\bibfnamefont {M.}~\bibnamefont
			{Vojta}}, \bibinfo {author} {\bibfnamefont {R.}~\bibnamefont {Bulla}},\ and\
		\bibinfo {author} {\bibfnamefont {W.}~\bibnamefont {Hofstetter}},\ }\bibfield
	{title} {\bibinfo {title} {Quantum phase transitions in models of coupled
			magnetic impurities},\ }\href@noop {} {\bibfield  {journal} {\bibinfo
			{journal} {Phys. Rev. B}\ }\textbf {\bibinfo {volume} {65}},\ \bibinfo
		{pages} {140405} (\bibinfo {year} {2002})}\BibitemShut {NoStop}%
	\bibitem [{\citenamefont {Borda}\ \emph {et~al.}(2003)\citenamefont {Borda},
		\citenamefont {Zarand}, \citenamefont {Hofstetter}, \citenamefont
		{Halperin},\ and\ \citenamefont {von Delft}}]{bordaPRL03}%
	\BibitemOpen
	\bibfield  {author} {\bibinfo {author} {\bibfnamefont {L.}~\bibnamefont
			{Borda}}, \bibinfo {author} {\bibfnamefont {G.}~\bibnamefont {Zarand}},
		\bibinfo {author} {\bibfnamefont {W.}~\bibnamefont {Hofstetter}}, \bibinfo
		{author} {\bibfnamefont {B.~I.}\ \bibnamefont {Halperin}},\ and\ \bibinfo
		{author} {\bibfnamefont {J.}~\bibnamefont {von Delft}},\ }\bibfield  {title}
	{\bibinfo {title} {Su(4) fermi liquid state and spin filtering in a double
			quantum dot system},\ }\href@noop {} {\bibfield  {journal} {\bibinfo
			{journal} {Phys. Rev. Lett.}\ }\textbf {\bibinfo {volume} {90}},\ \bibinfo
		{pages} {026602} (\bibinfo {year} {2003})}\BibitemShut {NoStop}%
	\bibitem [{\citenamefont {Craig}\ \emph {et~al.}(2004)\citenamefont {Craig},
		\citenamefont {Taylor}, \citenamefont {Lester}, \citenamefont {Marcus},
		\citenamefont {Hanson},\ and\ \citenamefont {Gossard}}]{Craig2004Science04}%
	\BibitemOpen
	\bibfield  {author} {\bibinfo {author} {\bibfnamefont {N.~J.}\ \bibnamefont
			{Craig}}, \bibinfo {author} {\bibfnamefont {J.~M.}\ \bibnamefont {Taylor}},
		\bibinfo {author} {\bibfnamefont {E.~A.}\ \bibnamefont {Lester}}, \bibinfo
		{author} {\bibfnamefont {C.~M.}\ \bibnamefont {Marcus}}, \bibinfo {author}
		{\bibfnamefont {M.~P.}\ \bibnamefont {Hanson}},\ and\ \bibinfo {author}
		{\bibfnamefont {A.~C.}\ \bibnamefont {Gossard}},\ }\bibfield  {title}
	{\bibinfo {title} {Tunable nonlocal spin control in a coupled-quantum dot
			system},\ }\href@noop {} {\bibfield  {journal} {\bibinfo  {journal}
			{Science}\ }\textbf {\bibinfo {volume} {304}},\ \bibinfo {pages} {565}
		(\bibinfo {year} {2004})}\BibitemShut {NoStop}%
	\bibitem [{\citenamefont {Sato}\ and\ \citenamefont {Eto}(2005)}]{SatoP05}%
	\BibitemOpen
	\bibfield  {author} {\bibinfo {author} {\bibfnamefont {T.}~\bibnamefont
			{Sato}}\ and\ \bibinfo {author} {\bibfnamefont {M.}~\bibnamefont {Eto}},\
	}\bibfield  {title} {\bibinfo {title} {Numerical renormalization group
			studies of su(4) kondo effect in quantum dots},\ }\href@noop {} {\bibfield
		{journal} {\bibinfo  {journal} {Physica E: Low-Dimensional Systems and
				Nanostructures}\ }\textbf {\bibinfo {volume} {29}},\ \bibinfo {pages} {652}
		(\bibinfo {year} {2005})}\BibitemShut {NoStop}%
	\bibitem [{\citenamefont {Galpin}\ \emph {et~al.}(2005)\citenamefont {Galpin},
		\citenamefont {Logan},\ and\ \citenamefont {Krishnamurthy}}]{GalpinPRL05}%
	\BibitemOpen
	\bibfield  {author} {\bibinfo {author} {\bibfnamefont {M.~R.}\ \bibnamefont
			{Galpin}}, \bibinfo {author} {\bibfnamefont {D.~E.}\ \bibnamefont {Logan}},\
		and\ \bibinfo {author} {\bibfnamefont {H.~R.}\ \bibnamefont
			{Krishnamurthy}},\ }\bibfield  {title} {\bibinfo {title} {Quantum phase
			transition in capacitively coupled double quantum dots},\ }\href@noop {}
	{\bibfield  {journal} {\bibinfo  {journal} {Phys. Rev. Lett.}\ }\textbf
		{\bibinfo {volume} {94}},\ \bibinfo {pages} {186406} (\bibinfo {year}
		{2005})}\BibitemShut {NoStop}%
	\bibitem [{\citenamefont {Ruiz-Tijerina}\ \emph {et~al.}(2014)\citenamefont
		{Ruiz-Tijerina}, \citenamefont {Vernek},\ and\ \citenamefont
		{Ulloa}}]{RuizPRB14}%
	\BibitemOpen
	\bibfield  {author} {\bibinfo {author} {\bibfnamefont {D.~A.}\ \bibnamefont
			{Ruiz-Tijerina}}, \bibinfo {author} {\bibfnamefont {E.}~\bibnamefont
			{Vernek}},\ and\ \bibinfo {author} {\bibfnamefont {S.~E.}\ \bibnamefont
			{Ulloa}},\ }\bibfield  {title} {\bibinfo {title} {Capacitive interactions and
			kondo effect tuning in double quantum impurity systems},\ }\href@noop {}
	{\bibfield  {journal} {\bibinfo  {journal} {Phys. Rev. B}\ }\textbf {\bibinfo
			{volume} {90}},\ \bibinfo {pages} {035119} (\bibinfo {year}
		{2014})}\BibitemShut {NoStop}%
	\bibitem [{\citenamefont {Nishikawa}\ \emph {et~al.}(2016)\citenamefont
		{Nishikawa}, \citenamefont {Curtin}, \citenamefont {Hewson}, \citenamefont
		{Crow},\ and\ \citenamefont {Bauer}}]{NishikawaPRB16}%
	\BibitemOpen
	\bibfield  {author} {\bibinfo {author} {\bibfnamefont {Y.}~\bibnamefont
			{Nishikawa}}, \bibinfo {author} {\bibfnamefont {O.~J.}\ \bibnamefont
			{Curtin}}, \bibinfo {author} {\bibfnamefont {A.~C.}\ \bibnamefont {Hewson}},
		\bibinfo {author} {\bibfnamefont {D.~J.~G.}\ \bibnamefont {Crow}},\ and\
		\bibinfo {author} {\bibfnamefont {J.}~\bibnamefont {Bauer}},\ }\bibfield
	{title} {\bibinfo {title} {Conditions for observing emergent su(4) symmetry
			in a double quantum dot},\ }\href@noop {} {\bibfield  {journal} {\bibinfo
			{journal} {Phys. Rev. B}\ }\textbf {\bibinfo {volume} {93}},\ \bibinfo
		{pages} {235115} (\bibinfo {year} {2016})}\BibitemShut {NoStop}%
	\bibitem [{\citenamefont {Lopes}\ \emph {et~al.}(2017)\citenamefont {Lopes},
		\citenamefont {Padilla}, \citenamefont {Martins},\ and\ \citenamefont
		{Anda}}]{Lopes2017Jun}%
	\BibitemOpen
	\bibfield  {author} {\bibinfo {author} {\bibfnamefont {V.}~\bibnamefont
			{Lopes}}, \bibinfo {author} {\bibfnamefont {R.~A.}\ \bibnamefont {Padilla}},
		\bibinfo {author} {\bibfnamefont {G.~B.}\ \bibnamefont {Martins}},\ and\
		\bibinfo {author} {\bibfnamefont {E.~V.}\ \bibnamefont {Anda}},\ }\bibfield
	{title} {\bibinfo {title} {{SU(4)-SU(2) crossover and spin-filter properties
				of a double quantum dot nanosystem}},\ }\href
	{https://doi.org/10.1103/PhysRevB.95.245133} {\bibfield  {journal} {\bibinfo
			{journal} {Phys. Rev. B}\ }\textbf {\bibinfo {volume} {95}},\ \bibinfo
		{pages} {245133} (\bibinfo {year} {2017})}\BibitemShut {NoStop}%
	\bibitem [{\citenamefont {Amasha}\ \emph {et~al.}(2013)\citenamefont {Amasha},
		\citenamefont {Keller}, \citenamefont {Rau}, \citenamefont {Carmi},
		\citenamefont {Katine}, \citenamefont {Shtrikman}, \citenamefont {Oreg},\
		and\ \citenamefont {Goldhaber-Gordon}}]{AmashaPRL13}%
	\BibitemOpen
	\bibfield  {author} {\bibinfo {author} {\bibfnamefont {S.}~\bibnamefont
			{Amasha}}, \bibinfo {author} {\bibfnamefont {A.~J.}\ \bibnamefont {Keller}},
		\bibinfo {author} {\bibfnamefont {I.~G.}\ \bibnamefont {Rau}}, \bibinfo
		{author} {\bibfnamefont {A.}~\bibnamefont {Carmi}}, \bibinfo {author}
		{\bibfnamefont {J.~A.}\ \bibnamefont {Katine}}, \bibinfo {author}
		{\bibfnamefont {H.}~\bibnamefont {Shtrikman}}, \bibinfo {author}
		{\bibfnamefont {Y.}~\bibnamefont {Oreg}},\ and\ \bibinfo {author}
		{\bibfnamefont {D.}~\bibnamefont {Goldhaber-Gordon}},\ }\bibfield  {title}
	{\bibinfo {title} {Pseudospin-resolved transport spectroscopy of the kondo
			effect in a double quantum dot},\ }\href@noop {} {\bibfield  {journal}
		{\bibinfo  {journal} {Phys. Rev. Lett.}\ }\textbf {\bibinfo {volume} {110}},\
		\bibinfo {pages} {046604} (\bibinfo {year} {2013})}\BibitemShut {NoStop}%
	\bibitem [{\citenamefont {Keller}\ \emph {et~al.}(2014)\citenamefont {Keller},
		\citenamefont {Amasha}, \citenamefont {Weymann}, \citenamefont {Moca},
		\citenamefont {Rau}, \citenamefont {Katine}, \citenamefont {Shtrikman},
		\citenamefont {Zar{\ifmmode\acute{a}\else\'{a}\fi}nd},\ and\ \citenamefont
		{Goldhaber-Gordon}}]{kellerNP14}%
	\BibitemOpen
	\bibfield  {author} {\bibinfo {author} {\bibfnamefont {A.~J.}\ \bibnamefont
			{Keller}}, \bibinfo {author} {\bibfnamefont {S.}~\bibnamefont {Amasha}},
		\bibinfo {author} {\bibfnamefont {I.}~\bibnamefont {Weymann}}, \bibinfo
		{author} {\bibfnamefont {C.~P.}\ \bibnamefont {Moca}}, \bibinfo {author}
		{\bibfnamefont {I.~G.}\ \bibnamefont {Rau}}, \bibinfo {author} {\bibfnamefont
			{J.~A.}\ \bibnamefont {Katine}}, \bibinfo {author} {\bibfnamefont
			{H.}~\bibnamefont {Shtrikman}}, \bibinfo {author} {\bibfnamefont
			{G.}~\bibnamefont {Zar{\ifmmode\acute{a}\else\'{a}\fi}nd}},\ and\ \bibinfo
		{author} {\bibfnamefont {D.}~\bibnamefont {Goldhaber-Gordon}},\ }\bibfield
	{title} {\bibinfo {title} {{Emergent SU(4) Kondo physics in a
				spin{\textendash}charge-entangled double quantum dot}},\ }\href
	{https://doi.org/10.1038/nphys2844} {\bibfield  {journal} {\bibinfo
			{journal} {Nat. Phys.}\ }\textbf {\bibinfo {volume} {10}},\ \bibinfo {pages}
		{145} (\bibinfo {year} {2014})}\BibitemShut {NoStop}%
	\bibitem [{\citenamefont {Leijnse}\ and\ \citenamefont
		{Flensberg}(2012)}]{leijnse_ParityQubits_2012}%
	\BibitemOpen
	\bibfield  {author} {\bibinfo {author} {\bibfnamefont {M.}~\bibnamefont
			{Leijnse}}\ and\ \bibinfo {author} {\bibfnamefont {K.}~\bibnamefont
			{Flensberg}},\ }\bibfield  {title} {\bibinfo {title} {Parity qubits and poor
			man's {{Majorana}} bound states in double quantum dots},\ }\href
	{https://doi.org/10.1103/PhysRevB.86.134528} {\bibfield  {journal} {\bibinfo
			{journal} {Phys. Rev. B}\ }\textbf {\bibinfo {volume} {86}},\ \bibinfo
		{pages} {134528} (\bibinfo {year} {2012})}\BibitemShut {NoStop}%
	\bibitem [{\citenamefont {Ivanov}(2017)}]{Ivanov2017Jul}%
	\BibitemOpen
	\bibfield  {author} {\bibinfo {author} {\bibfnamefont {T.~I.}\ \bibnamefont
			{Ivanov}},\ }\bibfield  {title} {\bibinfo {title} {{Coherent tunneling
				through a double quantum dot coupled to Majorana bound states}},\ }\href
	{https://doi.org/10.1103/PhysRevB.96.035417} {\bibfield  {journal} {\bibinfo
			{journal} {Phys. Rev. B}\ }\textbf {\bibinfo {volume} {96}},\ \bibinfo
		{pages} {035417} (\bibinfo {year} {2017})}\BibitemShut {NoStop}%
	\bibitem [{\citenamefont {Cifuentes}\ and\ \citenamefont
		{da~Silva}(2019)}]{cifuentes_ManipulatingMajorana_2019}%
	\BibitemOpen
	\bibfield  {author} {\bibinfo {author} {\bibfnamefont {J.~D.}\ \bibnamefont
			{Cifuentes}}\ and\ \bibinfo {author} {\bibfnamefont {L.~G. G. V.~D.}\
			\bibnamefont {da~Silva}},\ }\bibfield  {title} {\bibinfo {title}
		{Manipulating {{Majorana}} zero modes in double quantum dots},\ }\href
	{https://doi.org/10.1103/PhysRevB.100.085429} {\bibfield  {journal} {\bibinfo
			{journal} {Phys. Rev. B}\ }\textbf {\bibinfo {volume} {100}},\ \bibinfo
		{pages} {085429} (\bibinfo {year} {2019})}\BibitemShut {NoStop}%
	\bibitem [{\citenamefont {Ran{\v c}i{\'c}}\ \emph {et~al.}(2019)\citenamefont
		{Ran{\v c}i{\'c}}, \citenamefont {Hoffman}, \citenamefont {Schrade},
		\citenamefont {Klinovaja},\ and\ \citenamefont
		{Loss}}]{rancic_EntanglingSpins_2019}%
	\BibitemOpen
	\bibfield  {author} {\bibinfo {author} {\bibfnamefont {M.~J.}\ \bibnamefont
			{Ran{\v c}i{\'c}}}, \bibinfo {author} {\bibfnamefont {S.}~\bibnamefont
			{Hoffman}}, \bibinfo {author} {\bibfnamefont {C.}~\bibnamefont {Schrade}},
		\bibinfo {author} {\bibfnamefont {J.}~\bibnamefont {Klinovaja}},\ and\
		\bibinfo {author} {\bibfnamefont {D.}~\bibnamefont {Loss}},\ }\bibfield
	{title} {\bibinfo {title} {Entangling spins in double quantum dots and
			{{Majorana}} bound states},\ }\href
	{https://doi.org/10.1103/PhysRevB.99.165306} {\bibfield  {journal} {\bibinfo
			{journal} {Phys. Rev. B}\ }\textbf {\bibinfo {volume} {99}},\ \bibinfo
		{pages} {165306} (\bibinfo {year} {2019})}\BibitemShut {NoStop}%
	\bibitem [{\citenamefont {Chen}\ \emph {et~al.}(2020)\citenamefont {Chen},
		\citenamefont {Feng},\ and\ \citenamefont {Wang}}]{Chen2020Jul}%
	\BibitemOpen
	\bibfield  {author} {\bibinfo {author} {\bibfnamefont {Y.-A.}\ \bibnamefont
			{Chen}}, \bibinfo {author} {\bibfnamefont {J.-J.}\ \bibnamefont {Feng}},\
		and\ \bibinfo {author} {\bibfnamefont {Z.}~\bibnamefont {Wang}},\ }\bibfield
	{title} {\bibinfo {title} {{Proposal for probing the Majorana zero modes by
				testing the Pauli exclusion principle with two quantum dots}},\ }\href
	{https://doi.org/10.1016/j.physleta.2020.126496} {\bibfield  {journal}
		{\bibinfo  {journal} {Phys. Lett. A}\ }\textbf {\bibinfo {volume} {384}},\
		\bibinfo {pages} {126496} (\bibinfo {year} {2020})}\BibitemShut {NoStop}%
	\bibitem [{\citenamefont {Wilson}(1975)}]{WilsonRMP75}%
	\BibitemOpen
	\bibfield  {author} {\bibinfo {author} {\bibfnamefont {K.~G.}\ \bibnamefont
			{Wilson}},\ }\bibfield  {title} {\bibinfo {title} {The renormalization group:
			Critical phenomena and the kondo problem},\ }\href@noop {} {\bibfield
		{journal} {\bibinfo  {journal} {Rev. Mod. Phys.}\ }\textbf {\bibinfo {volume}
			{47}},\ \bibinfo {pages} {773} (\bibinfo {year} {1975})}\BibitemShut
	{NoStop}%
	\bibitem [{\citenamefont {Flensberg}(2010)}]{Flensberg2010Nov}%
	\BibitemOpen
	\bibfield  {author} {\bibinfo {author} {\bibfnamefont {K.}~\bibnamefont
			{Flensberg}},\ }\bibfield  {title} {\bibinfo {title} {{Tunneling
				characteristics of a chain of Majorana bound states}},\ }\href
	{https://doi.org/10.1103/PhysRevB.82.180516} {\bibfield  {journal} {\bibinfo
			{journal} {Phys. Rev. B}\ }\textbf {\bibinfo {volume} {82}},\ \bibinfo
		{pages} {180516(R)} (\bibinfo {year} {2010})}\BibitemShut {NoStop}%
	\bibitem [{\citenamefont {Bulla}\ \emph {et~al.}(2008)\citenamefont {Bulla},
		\citenamefont {Costi},\ and\ \citenamefont {Pruschke}}]{BullaRMP08}%
	\BibitemOpen
	\bibfield  {author} {\bibinfo {author} {\bibfnamefont {R.}~\bibnamefont
			{Bulla}}, \bibinfo {author} {\bibfnamefont {T.~A.}\ \bibnamefont {Costi}},\
		and\ \bibinfo {author} {\bibfnamefont {T.}~\bibnamefont {Pruschke}},\
	}\bibfield  {title} {\bibinfo {title} {Numerical renormalization group method
			for quantum impurity systems},\ }\href@noop {} {\bibfield  {journal}
		{\bibinfo  {journal} {Rev. Mod. Phys.}\ }\textbf {\bibinfo {volume} {80}},\
		\bibinfo {pages} {395} (\bibinfo {year} {2008})}\BibitemShut {NoStop}%
	\bibitem [{Fle()}]{FlexibleDMNRG}%
	\BibitemOpen
	\href@noop {} {\bibinfo  {journal} {We used the open-access Budapest Flexible
			DM-NRG code, http://www.phy.bme.hu/\~{}dmnrg/; O. Legeza, C. P. Moca, A. I.
			T\'{o}th, I. Weymann, G. Zar\'{a}nd, arXiv:0809.3143 (2008) (unpublished)}\
	}\BibitemShut {NoStop}%
	\bibitem [{\citenamefont {Weymann}(2011)}]{weymannPRB11}%
	\BibitemOpen
	\bibfield  {journal} {  }\bibfield  {author} {\bibinfo {author} {\bibfnamefont
			{I.}~\bibnamefont {Weymann}},\ }\bibfield  {title} {\bibinfo {title}
		{Finite-temperature spintronic transport through kondo quantum dots:
			Numerical renormalization group study},\ }\href@noop {} {\bibfield  {journal}
		{\bibinfo  {journal} {Phys. Rev. B}\ }\textbf {\bibinfo {volume} {83}},\
		\bibinfo {pages} {113306} (\bibinfo {year} {2011})}\BibitemShut {NoStop}%
\end{thebibliography}

%

\end{document}